\documentclass[mnsc,noblindrev]{informs3}

\OneAndAHalfSpacedXII 

\usepackage{mathtools}
\usepackage{amsmath}
\usepackage{color,soul}
\usepackage{bm}
\usepackage{cleveref, comment}
\usepackage{subcaption}
\usepackage{enumerate}
\usepackage[section]{placeins}
\usepackage{array}

\newsavebox{\hiddenbox}
\newcommand{\ZZ}{\hspace{\tabcolsep}}
\newcolumntype{Z}{>{\begin{lrbox}{\hiddenbox}}p{0pt}<{\end{lrbox}}}
\newcolumntype{U}{@{\ZZ}Z@{}}
\newcolumntype{V}{@{\ZZ}Z@{\ZZ}}

\usepackage{natbib}
 \bibpunct[, ]{(}{)}{,}{a}{}{,}%
 \def\bibfont{\small}%

\newcommand\T{\rule{0pt}{2.6ex}}       

\newcommand\Panda{Top Handcrafted Features}       
\newcommand\TempoF{Tempo Features}

\TheoremsNumberedThrough     
\ECRepeatTheorems

\EquationsNumberedThrough    

\MANUSCRIPTNO{}

\RRHFirstLine{}
\RRHSecondLine{Explainable Deep Learning for Music Emotion}

\LRHFirstLine{}
\LRHSecondLine{Explainable Deep Learning for Music Emotion}

\begin{document}


\RUNAUTHOR{Fong, Kumar, and Sudhir}

\RUNTITLE{Explainable Deep Learning for Music Emotion}

\TITLE{A Theory-Based Explainable Deep Learning \\Architecture for Music Emotion}

\ARTICLEAUTHORS{%
\AUTHOR{Hortense Fong}
\AFF{Columbia Business School, \EMAIL{hf2462@gsb.columbia.edu}} 
\AUTHOR{Vineet Kumar, K. Sudhir}
\AFF{Yale School of Management, \EMAIL{vineet.kumar@yale.edu}, \EMAIL{k.sudhir@yale.edu}} 
} 
\JOURNAL{Marketing Science}

\ABSTRACT{This paper develops a theory-based, explainable deep learning convolutional neural network (CNN) classifier to predict the time-varying emotional response to music. We design novel CNN filters that leverage the frequency harmonics structure from acoustic physics known to impact the perception of musical features. Our theory-based model is more parsimonious, but provides comparable predictive performance to atheoretical deep learning models, while performing better than models using handcrafted features. Our model can be complemented with handcrafted features, but the performance improvement is marginal. Importantly, the harmonics-based structure placed on the CNN filters provides better explainability for how the model predicts emotional response (valence and arousal), because emotion is closely related to consonance---a perceptual feature defined by the alignment of harmonics. Finally, we illustrate the utility of our model with an application involving digital advertising. Motivated by YouTube’s mid-roll ads, we conduct a lab experiment in which we exogenously insert ads at different times within videos. We find that ads placed in emotionally similar contexts increase ad engagement (lower skip rates, higher brand recall rates). Ad insertion based on emotional similarity metrics predicted by our theory-based, explainable model produces comparable or better engagement relative to atheoretical models.}

\maketitle



\section{Introduction}\label{section:introduction}

Music is widely regarded as among the most effective and efficient of channels to influence emotion; it is often called the \emph{language of emotion} \citep{corrigall2013music}. As emotions play a central role in many elements of marketing and consumer behavior, marketers routinely use music to evoke emotion along the customer journey, from need recognition to purchase, in advertising, content marketing, and stores \citep{gorn1982effects,krishna2012integrative}. In particular, music is almost universally used in advertising and advertisers spend considerable effort crafting it to elicit desired emotions and consequent marketing outcomes.\footnote{A content analysis of over 3,000 ads showed that 94\% of ads use music \citep{allan2008content}. Further, over 75\% of advertising hours in broadcast media uses music in some form \citep{huron1989music}. As \cite{huron1989music} states: ``on a second-for-second basis, advertising music is the most meticulously crafted and most fretted about music…’’} As such, a tool that maps music (or an ad with music) to its evoked emotion can be valuable to marketers. For example, marketers can use such a method to automate emotion-based contextual matching of ads and content at scale to increase ad engagement. On music and video platforms, it can be valuable in automating creation of mood-based playlists and ``next song'' recommendations for users from their large collections.\footnote{As of 2024, YouTube has 2.49 billion users, spending on average 49 minutes per day on YouTube. Spotify has 615 million users whose playlists span over 100 million tracks.}

In this paper, we develop a theory-based, explainable deep learning convolutional neural network (CNN) classifier that takes music audio as input and predicts the sequence of emotions it evokes in a listener. We characterize emotion using the well-established valence-arousal framework developed by \cite{russell1980circumplex}. Valence measures how positive or negative a listener feels and higher valence maps to more positive feeling. Arousal measures how energetic a listener feels and higher arousal maps to greater excitement and energy.\footnote{Many researchers have adopted the valence-arousal framework for music emotion classification (e.g., \citealt{panda2018novel, yang2011predicting}). Our model can also easily be adapted for other emotion frameworks, such as the discrete emotion framework.}

Existing music emotion classifiers can be grouped into two types based on explainability and accuracy, and there is often a tradeoff between these. Classifiers using handcrafted features and traditional machine learning methods are typically more explainable, but often less accurate whereas classifiers using data-driven variables and ``atheoretical" deep learning methods are typically less explainable but more accurate. By incorporating theory into the model design, our deep learning model seeks to be both accurate and explainable. 

We highlight and explain the two key distinguishing features in the modeling contribution. First, and most important, we incorporate music theory based on harmonics in the model. When a musical note is generated, its pitch is characterized by its fundamental frequency (i.e., lowest frequency) and harmonics refer to the whole number multiples of the fundamental frequency. We \emph{design harmonics filters} to capture theoretically known musical features that link to emotion.\footnote{The theory-based filters are distinct from handcrafted features (e.g., MFCCs) that are used with traditional machine learning models (e.g., support vector machine) because the filter operations are learned from the data rather than predefined.} Such filter construction is challenging because current CNN implementations for music are based on adaptations of models developed for computer vision where \emph{spatial contiguity} is meaningful. In music, however, many important features are dependent on elements that are \emph{non-contiguous}, implying that filters inspired by vision are inadequate and/or inefficient for music. Harmonics, in particular, are by definition based on non-contiguous frequencies and underlie several musical features, like consonance, timbre, and pitch, which have well-established empirical links to emotional response.\footnote{See Appendix Table \ref{table:music_definitions} for music definitions.} We therefore incorporate structure into the CNN filters based on the role of harmonics, which also helps to create a more parsimonious model without reducing accuracy, relative to benchmark atheoretical models.

Second, our theory-based filters are more explainable, relative to atheoretical deep learning models, because the filters are designed to characterize features that are well-known to impact emotion. To aid explainability, we visualize the link between emotion and the features learned by the theory-based filters. For this, we adapt recent advances in the visualization of deep learning models for computer vision (i.e., Grad-CAM). The focus on explainability allows us to go beyond model prediction, giving researchers and practitioners confidence that the model is learning a feature of the music (i.e., consonance) that can generalize outside of the training data rather than picking up spurious correlations. Explainability helps build trust in the model, which facilitates adoption at scale.\footnote{In 2020, McKinsey conducted a survey that found that explainability is among the top risks firms are concerned about regarding AI models. Source: \url{https://www.mckinsey.com/business-functions/quantumblack/our-insights/global-survey-the-state-of-ai-in-2020}} 

Finally, we illustrate the practical value of our model in an application motivated by YouTube's mid-roll video ads. YouTube serves tens of millions of  content videos each day; an ad can be placed in a number of different ad breaks within each content video. Given the scale, identifying the optimal ad breaks in the videos for ad insertion requires automation. First, we examine whether matching the ad's emotion with the emotion of the content at the ad insertion point improves or hurts ad effectiveness. To test this, we conduct a lab experiment in which we insert ads at points in videos that differ in emotional similarity between the ads and content videos. We find that greater emotional similarity helps by decreasing ad skip and increasing brand recall. We then input the audio of the ads and content videos into our model to predict emotion and use the predicted emotion to determine the most emotionally similar ad insertion point for each ad within each content video.\footnote{We use the emotion elicited by only the music as a proxy (or the dominant modality) for the emotion elicited by the video. This is reasonable given that music is typically designed to elicit the intended emotion in videos \citep{bullerjahn1994empirical,herget2021music}. We also examine how other modalities in video (e.g., text of dialogues, images) can be used alongside music emotion in \S \ref{section:other_modalities}.} The engagement outcomes based on ad insertion using our theory-driven model are comparable to those from ad insertion using atheoretical models, but unlike the atheoretical models, our model is more explainable. 

We provide an overview of the elements of our model to help build intuition for the critical challenges in constructing a CNN model for predicting music emotion that is both theory-based and explainable. Our methodological approach uses the raw sound wave of a music clip rather than pre-defined handcrafted features as the starting input. We transform the sound wave into a mel spectrogram, which reflects human hearing and is the input into the CNN. The mel spectrogram visualizes which frequencies are present at any given time.\footnote{More specifically, it is a two-dimensional image in which the $y$-axis denotes frequency, the $x$-axis denotes time, and color captures the magnitude of each frequency at each time point.} This representation is useful because it allows us to ``read" some of the clip's music features, such as frequency range and note density.  Recent research developing and applying CNN for music uses mel spectrograms as input (e.g., \citealt{pons2016music,chowdhury2019towards,rajaram2020video}), and we detail the rationale and implementation in \S \ref{section:model}.

CNNs are deep neural networks specially developed for image processing, in which objects are contiguous across both $x$- and $y$-dimensions, which have spatial meaning based on physical reality. Convolution filters play a critical role in determining the performance of CNNs, and the filters designed for image processing take advantage of spatial contiguity to perform effectively. However, spectrograms generated from music are not like regular images. In spectrograms, non-contiguous regions in the frequency space impact many characteristics of music that humans perceive. For example, an octave (e.g., A4 (440 Hz) and A5 (880 Hz)) produces a consonant sound while a tritone (e.g., A4 (440 Hz) and D5 (587 Hz)) produces a dissonant sound. Typical square CNN filters, which account for contiguous areas of an image, cannot capture such concepts based on non-contiguous frequencies, thus highlighting the challenge of designing novel filters to incorporate music domain knowledge into deep learning models.\footnote{\cite{goodfellow2016deep} write in their textbook the following about square convolution filters: ``When a task involves incorporating information from very distant locations in the input, then the prior imposed by convolution may be inappropriate."} 

We develop novel non-contiguous filters that specifically incorporate relevant harmonic frequencies into the CNN model. By incorporating the harmonic structure, the filters can more parsimoniously capture many relevant characteristics of music that involve harmonics and their interactions. In particular, they help to identify consonance, which is well-known to influence emotional response to music.\footnote{Consonance refers to a combination of notes that sound pleasant when played simultaneously. Dissonance, or the lack of consonance, refers to a combination of notes that sound jarring when played simultaneously \citep{muller2015fundamentals}. Consonance is one of the few features that has a strong relationship with valence. Models that predict music emotion have historically been better at classifying arousal and worse at classifying valence \citep{aljanaki2017developing}.} By leveraging ideas from music theory to build the model, we set the stage for explainability post-estimation.

To aid explainability, we adapt Grad-CAM, a tool that uses a heatmap to visualize which areas of an image contribute most to the classification of the image into a specific class \citep{selvaraju2017grad}. Consider the problem of classifying images of wolves and huskies. Suppose a CNN trained to distinguish wolves from huskies generates Grad-CAM heatmaps that highlight the background of the image as the main predictor because wolves are commonly photographed in snow and huskies in grass \citep{ribeiro2016should}. If the prediction is driven by the background snow or grass, a spurious correlation, a new example of a husky playing in snow is more likely to be classified incorrectly as a wolf. With Grad-CAM highlighting the background, we learn that the model did not learn to distinguish between wolves and huskies based on features of the animals themselves, but rather the background of the image, implying that the generalizability of the model is under question. Knowing \emph{why} a model makes various predictions helps assess model generalizability.

In the context of music, relating the top-level outcome or label of interest (e.g., emotion) to a mid-level set of musical features (e.g., consonance, timbre, rhythm) provides us a clearer understanding of the features responsible for the prediction, making the model more explainable.\footnote{Appendix \Cref{table:music_feature_interpretability} organizes features used in music classification by level of interpretability.} While low-level features (e.g., frequency, time) provide some degree of transparency,   they do not have a clear explainable link to top-level labels \citep{fu2010survey}. Our Grad-CAM visualizations for the theory-based deep learning model highlight areas of high and low consonance in the mel spectrograms consistent with theory linking consonance and emotion. In contrast, heatmaps for the model based on atheoretical square filters show a less clear pattern linking musical features to emotion. Broadly, this contrast shows the value of incorporating theory-based structure into CNN filter design for explainability.

Summarizing, our key contributions are as follows. First, we develop a theory-based, explainable deep learning framework that models and predicts time-varying emotional response. Second, our approach of integrating a harmonics-based structure enables model explainability. We show with Grad-CAM that incorporating such structure into the design of the CNN filter allows us to visualize the link between predicted emotion and consonance. Third, the theory-based model produces comparable predictive performance to atheoretical models with much greater parsimony, which in turn makes the model less complex to train, as well as more  generalizable and robust (e.g., \citealt{gerg2021structural,kutz2022parsimony}). Fourth, we show how additional handcrafted features can be incorporated into our deep learning framework without impacting explainability. Fifth, we demonstrate the model's practical value with an application that examines the impact of matching the emotion of a video ad with the emotion of the content video.  

\section{Related Literature}\label{section:literature}

Our paper builds upon several distinct streams of literature across fields, as detailed below. 

\subsection{Listener Response to Music}

Music induces emotion, as shown by a wide literature using methods ranging from surveys to brain scans \citep{johnson2016handbook}. Since our focus is on the background music of content and ads, which typically falls under Western tonal music \citep{stoppe2014film,nelson2013hollywood}, our focus in this paper is on the musical associations between Western tonal music and emotion.\footnote{\cite{nelson2013hollywood} writes ``Most films, however, are targeted at a broad, global audience with the implicit understanding that they share a common familiarity with Western tonal music, so the overtone series is the foundation of the cinematic musical language." According to \cite{christensen2006cambridge}, ``tonality most often refers to the orientation of melodies and harmonies toward a referential (or tonic) pitch class."} In particular, consonance and dissonance play a major role in creating emotion in background music. According to \cite{nelson2013hollywood}, the manipulation of consonant and dissonant harmonies ``is the most immediately effective way a composer can alter the mood of a scene." In Western tonal music, consonance is associated with positive valence and low arousal emotion, while dissonance is associated with negative valence and high arousal emotion \citep{gabrielsson2010role,thompson2010cross}. Appendix \ref{section:harmonics_and_features} discusses how timbre and pitch also impact emotional response. 

Emotional responses to music have marketing implications, and a substream of literature focuses on the relationship between music features and marketing outcomes. \cite{bruner1990music} overviews how music elicits different moods, which in turn impact ad outcomes. \cite{yang2022high} use low-level acoustic features to predict ad audio energy levels and find that energetic commercials are more likely to be watched for longer. \cite{boughanmi2021dynamics} use a Bayesian nonparametric approach to predict album sales using multi-modal data that includes high-level audio features of music in the songs of the albums. \cite{melzner2023sound} study how timbre affects judgments of brand personality. More broadly, our research is related to the literature on sensory marketing as music impacts consumers through the auditory sense \citep{krishna2012integrative}.

\subsection{Machine Learning with Unstructured Audio Data}

While audio includes both speech and music, our focus in this paper is on music. Traditional machine learning methods, like SVM, previously produced good classification performance in many settings by using handcrafted features, such as mel frequency cepstral coefficients (MFCCs). However, the performance of deep learning algorithms has overtaken almost all other methods in audio applications, similar to vision applications \citep{hinton2012deep}. The crucial advantage that deep learning has is that features are automatically learned from data, rather than predefined \citep{choi2017tutorial}. 

With deep learning, music is typically converted to a spectrogram, which is used as the input to the learning algorithm. A few researchers have attempted to build music-specific deep learning models rather than use CNN models designed for image recognition. For example, to predict ballroom music genre, \cite{pons2016music} suggest using musically-motivated CNN filters to capture low-level timbral and temporal elements of music. This translates to using various rectangular convolutional filters---tall and skinny filters for timbral elements and short and wide filters for temporal elements. Using data with labeled emotion and mid-level features (e.g., melodiousness, articulation), \cite{chowdhury2019towards} build a deep learning model that includes an interpretable mid-level feature layer to predict emotion. We contribute to the audio machine learning literature by designing and developing novel CNN filters based on harmonics, and demonstrate how the filters are useful for prediction as well as explanation. 

\subsection{Ad Insertion in Videos}\label{section:adinsertionlit}
An evolving literature has examined video ads within streaming videos to understand what ad, content video, and user characteristics impact ad performance. For a survey of this literature, see \cite{frade2021advertising}.\footnote{For a survey focused on television advertising, see \cite{wilbur2008two}.} The typical outcomes examined in studies of video ads include ad skipping, brand recognition, intrusiveness, aided and unaided brand recall, ad acceptance, click through rate, and related metrics for ads accompanied by a call to action. 

Research focused on ads has explored the impact of ad characteristics, such as length and content, and interactions between ads, such as the number of ads within a break, on outcomes like brand recall \citep{goodrich2015consumer}. Prior work exploring \textit{both ad and content video characteristics} has studied the impact of ad relevance (e.g., showing a Ford ad in a video about Formula 1 racing) and ad congruence on performance. For ad relevance, \cite{li2015you} found that for mid-roll ads, congruence between the ad product and the video content improves consumer receptivity, whereas the opposite is true for post-roll ads. For emotional congruence, results have been mixed. \cite{belanche2017understanding} found in an experiment that high arousal ads are watched for longer in high arousal content than in low arousal content but found no such effect of congruence for low arousal ads. On the other hand, although not in a video ad insertion context, \cite{puccinelli2015consumers} found that consumers in a low arousal state watch high arousal ads for less time than moderate-arousal ads, supporting congruence. \cite{kapoor2022does} focused on the emotions of happiness and sadness and found in a field experiment that emotional contrast led to greater ad engagement. Overall, the results in the literature are mixed on whether emotional congruence improves ad performance. An important point of contrast for this paper is that relative to past research  focusing on the overall emotion in the content video for ad matching, our focus is on the time-varying emotion of the content video to identify the optimal emotion-based ad insertion position within the video.

\section{Deep Learning Model for Music Emotion}\label{section:model}

We develop a CNN for emotion classification that incorporates theory relating to the physics of sound waves and the perception of Western tonal music by listeners. 

\subsection{Model Elements}\label{section:modelelements}
\begin{figure}[h]
    \centering
    \caption{Music Emotion Classification Schematic}
    \includegraphics[clip,width=\textwidth]{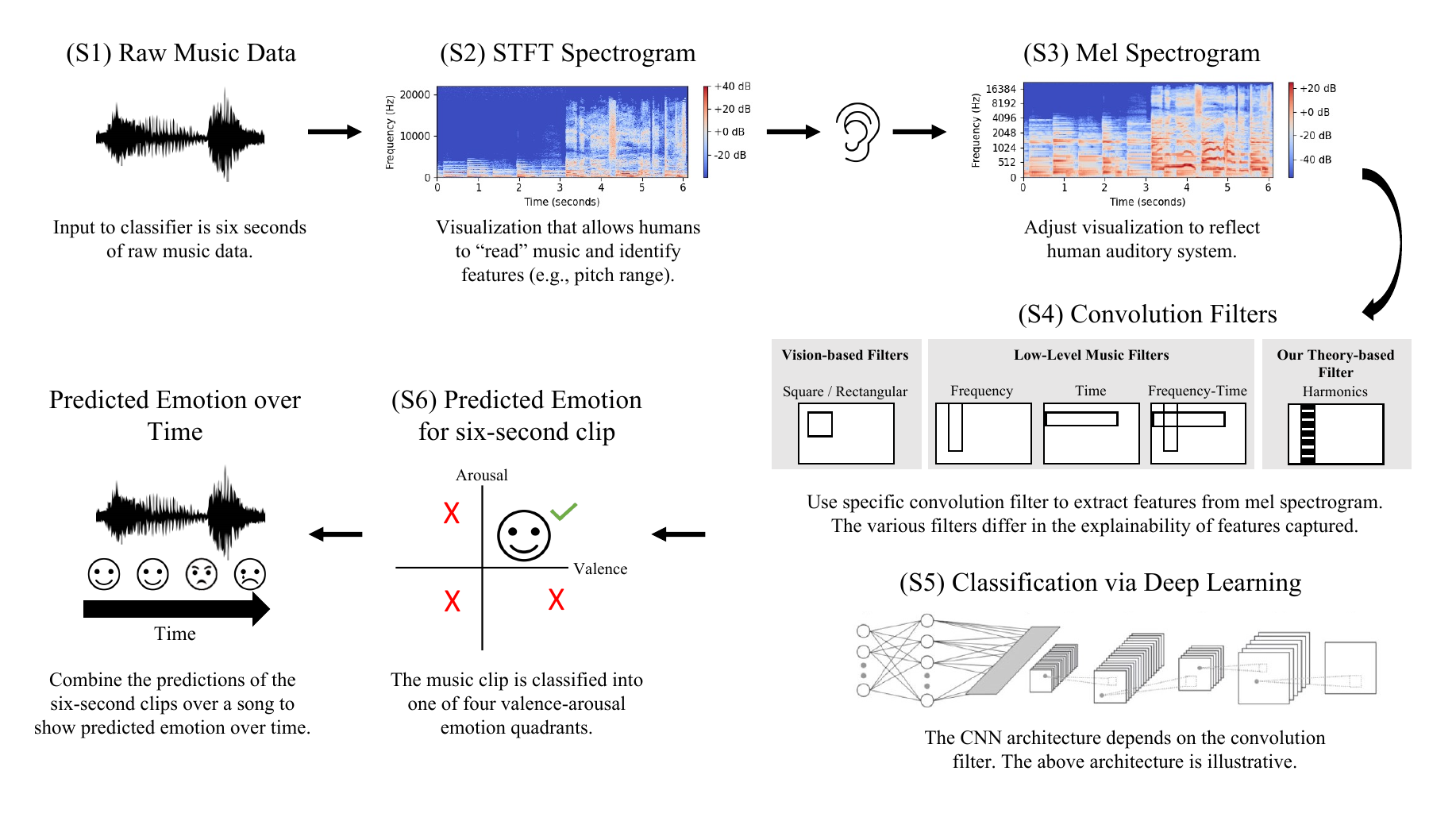}
    \label{figure:schematic}
\end{figure}

\Cref{figure:schematic} overviews the steps of our deep learning model that maps music to emotion. Step \textbf{S1} takes six seconds of raw audio sound wave data as input.\footnote{We use six seconds of audio but the model can easily be adapted to incorporate other audio lengths.} In Step \textbf{S2}, the music clip is converted to a short-time Fourier transform (STFT) spectrogram. In Step \textbf{S3}, we transform the STFT spectrogram to a mel spectrogram, which characterizes how the sound is perceived by the human ear. In Step \textbf{S4}, the mel spectrogram is used as a visual input to the CNN with one of the convolution filter types. We propose a theoretically motivated filter based on frequency harmonics to reflect aspects of music that we expect to impact listener emotion. For performance comparison, we also consider atheoretical square (and rectangular) filters that are commonly used in image processing as well as low-level music filters proposed in the literature. In Step \textbf{S5}, the features learned by the convolution filters from \textbf{S4} are put through the remaining layers of the CNN. In Step \textbf{S6}, the CNN  generates a classification prediction for the six-second sound clip, indicating the quadrant of the dimensional model into which the sound falls. Finally, combining the predictions of the  clips shows the emotional variation over time.

\paragraph{\textbf{(S1) Physical Properties of Sound Waves:}}
Music (or any sound) is a pressure wave that travels through the air until it reaches the listener's ear. The waveform illustrated in (S1) of \Cref{figure:schematic} graphs the change in air pressure at a certain location over six seconds \citep{muller2015fundamentals}. Audio data can be represented in a number of ways. While a waveform is one way to visually represent sound, it does not model how humans hear sound, which is based on the underlying frequencies.\footnote{Two different waveforms can map to the same sound.} 

\begin{figure}[h]
\caption{Example of Waveform and Spectrum}
\begin{minipage}{0.5\textwidth}
    \begin{center}
    \subcaption{\small Waveform}
    \includegraphics[width=0.8\textwidth]{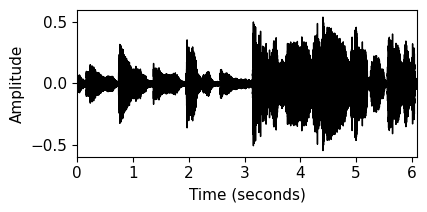}
    \label{figure:waveform}
    \end{center}
\end{minipage}
\begin{minipage}{0.5\textwidth}
    \begin{center}
    \subcaption{\small Spectrum}
    \includegraphics[width=0.8\textwidth]{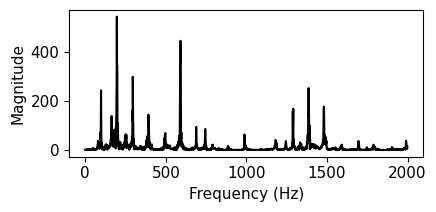}
    \label{figure:spectrum}
    \end{center}
\end{minipage}
\footnotesize
Notes: (a) Waveform of six seconds of New Soul by Yael Naim.
(b) Spectrum of the first second of the waveform.
\end{figure}

To get to musically relevant features, we need a representation of the different frequencies that the sound wave is composed of in terms of fundamental sine waves. Sine waves determine what humans hear and are at the foundation of musical concepts like pitch and harmony. The mathematical representation of this process is the Fourier transform, which decomposes a sound wave into its constituent sine waves. Any sound wave can be represented as a combination of sine waves of different frequencies, amplitudes, and phases, known as the \textit{partials} of the sound wave. The complete set of partials makes up the \textit{spectrum}. \Cref{figure:spectrum} shows the spectrum of the first second of music shown in the \Cref{figure:waveform} waveform. From the spectrum, we can identify the main frequencies that make up the sound and the magnitude of each frequency. 

\paragraph{\textbf{(S2) Short-Time Fourier Transform Spectrogram:}}
A spectrogram visualizes frequency and time features of audio data \citep{muller2015fundamentals}. The STFT spectrogram, which is produced by taking the Fourier transform of short overlapping time windows of the waveform to decompose it into its individual frequencies and their respective magnitudes, maps the squared magnitude of each frequency over time. The parameters that go into generating an STFT spectrogram are the sampling rate, window type and size, and hop length.\footnote{To operationalize this procedure, we first digitize the analog audio signal by sampling from the signal since we are working with digital technology. The sampling rate represents the number of samples taken per second and is measured in Hertz. The optimal sampling rate depends on the context. We will use a sampling rate of 44,100 Hz, which is also used for CD recordings, since it generates an STFT spectrogram that covers the range of human hearing, spanning from roughly 20 Hz to 20,000 Hz. Since time has been discretized, we now measure time in terms of number of samples rather than in seconds. We set the window type to Hann, the window size to 4,096 samples, and the hop length to 512 samples. These are standard choices in the literature \citep{muller2015fundamentals}. A Hann window is a bell-shaped window that places more weight on the center of the window and less weight on the edges of the window.} Let $x$ represent the discrete-time signal of the audio signal, $w$ the window function, which takes in $N$ samples, and $H$ the hop size. The window function specifies how we weight the audio signal within each window of time and the hop size specifies how many samples we jump between each window. The discrete STFT $X$ of signal $x$ is:
\begin{equation}
    X(m,k) \coloneqq \sum_{n=0}^{N-1} x(n + mH)w(n)exp(-2\pi ikn/N),
\end{equation}
where $m$ is the time index, $k \in [0:K]$ is the frequency index, and $i:=\sqrt{-1}$. A sampling rate of 44,100 Hz generates a spectrogram that extends up to 22,050 Hz. The STFT spectrogram can then be written as:
\begin{equation}
    S(m,k) \coloneqq |X(m,k)|^2.
\end{equation}
The magnitude of the complex number $X(m,k)$ captures the presence of each frequency at each time sample. Squaring the magnitude yields the power of each frequency at each time sample. We generate an STFT spectrogram for each six-second clip of music. The resulting frequency $\times$ time dimensions of the STFT spectrograms are $2049 \times 517$.\footnote{We follow the literature in choosing the dimensions of the spectrogram \citep{muller2015fundamentals}. The time dimension is equal to: length of music clip (6 sec) $\times$ sampling rate (44,100 Hz)/ hop length (512) =  $(6 \times 44100/512) = 517$. The frequency dimension is equal to: half the window sample size + 1 = $(4096 / 2 + 1) = 2049 $. The rationale for using half the window sample size is that in transforming the audio information from the time domain to the frequency domain, the discrete Fourier transform transforms the 4,096 samples to 4,096 frequency bins, but half of them are redundant. It is helpful to have high granularity on the frequency dimension since our theory-based convolution filters are focused on frequency. We observed empirically that using 4,096 frequency bins outperforms 2,048 frequency bins.}

\begin{figure}
\caption{Spectrograms}
\begin{minipage}{.5\textwidth}
    \begin{center}
    \subcaption{\small STFT Spectrogram}
    \includegraphics[width=0.95\textwidth]{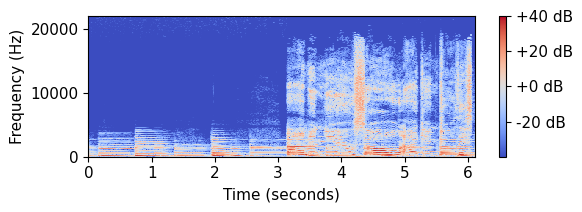}
    \label{figure:stft}
    \end{center}
\end{minipage}
\begin{minipage}{.5\textwidth}
    \begin{center}
    \subcaption{\small Mel Spectrogram}
    \includegraphics[width=0.95\textwidth]{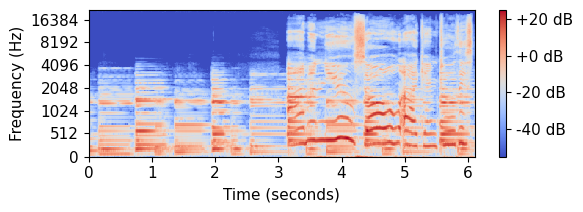}
    \label{figure:mel}
    \end{center}
\end{minipage}
\footnotesize
Notes: (a) STFT spectrogram of six seconds of New Soul shown in \Cref{figure:waveform}. The STFT spectrogram visualizes the time and frequency features of audio. The x-axis represents discretized time, the y-axis discretized frequency, and color the squared magnitude of each frequency bin over time. It enables one to ``read" musical features, such as the range of frequencies played.
(b) Mel spectrogram of six seconds of New Soul. The mel spectrogram transforms the linear frequency scale of the STFT to a log-frequency scale that reflects human auditory perception.
\end{figure}

The STFT spectrogram of \Cref{figure:waveform} is represented in \Cref{figure:stft}, with the $x$-dimension representing time, the $y$-dimension frequency, and color the power of each frequency bin at each time sample (red (blue) is high (low) power). The frequency and time dimensions are discretized since we are working with a digital signal. In the STFT spectrogram, frequency and time are shown on linear scales while power is shown on a log scale and measured in decibels (dB) since humans perceive volume on a log scale.\footnote{By using a log scale, small intensity values of relevance are visible to a human reader.} In \Cref{figure:stft}, the large patch of blue before second three indicates a lack of high frequencies early in the music.

\paragraph{\textbf{(S3) Mel Spectrogram based on Auditory Perception:}}
Humans are better at perceiving frequency differences at low pitches than at high pitches \citep{muller2015fundamentals}.\footnote{Pitch is a subjective measure of frequency and is defined as the attribute of sound that allows it to be ordered on a scale from low to high. For a pure tone sine wave, the pitch and frequency are the same, and are determined by its fundamental or lowest frequency. However, they can differ for more complex and realistic sounds. In all cases, the higher the frequency, the higher the perceived pitch.}
With equal sensitivity across the frequency spectrum, the STFT spectrogram does not represent human hearing. The mel spectrogram transforms the STFT spectrogram by mapping the frequencies onto the mel scale, a log-frequency scale created to reflect human hearing. Equal distances on the mel scale have the same perceptual distance in pitch.

The additional parameter that goes into generating a mel spectrogram from an STFT spectrogram is the number of mel bands, which specifies the mel filter banks---the weights that map the STFT frequencies to the mel frequency scale. We use 256 mel bands. The mel spectrogram in \Cref{figure:mel} more clearly displays the differences among the lower frequencies relative to \Cref{figure:stft}.

\paragraph{\textbf{(S4) Incorporating Harmonics-based Structure into CNN Filter Design:}}

Convolution filters in CNNs are matrix operations learned from the data with the goal of correctly making predictions. In designing convolution filters to learn features to classify music emotion, we must account for how humans perceive sound and in particular the non-contiguous structure of many musical features. We know from acoustic physics that harmonic frequencies underlie important mid-level musical features, like consonance, timbre, and pitch, which in turn relate to emotional response. We incorporate the harmonics-based structure into the CNN filters with two goals in mind: 1) relevant mid-level features are parsimoniously captured and 2) the model is more explainable relative to atheoretical deep learning models. We detail the concepts of harmonics and pitch class and how they connect to the development of theory-based filters next.

\subparagraph{Harmonics.}
The STFT from (S2) decomposes the sound wave into its constituent sine waves, known as partials. The harmonics of the sound wave are the partials that are integer multiples of its fundamental frequency (or lowest partial). Mathematically, the set of harmonics $\mathcal{H} \coloneqq \{\omega_n\}$ of a tone with fundamental frequency $f_0$ contains frequencies:
\begin{equation} \label{eq:harmonics}
\omega_n = n f_0 \quad \forall n \in \mathbb{Z}^+.
\end{equation}
\noindent Harmonics are the underlying reason for many musical features, and have been referred to as the ``fabric of music.''\footnote{Source: \url{https://splice.com/blog/what-are-harmonics/}} Appendix \ref{section:harmonics_and_features} provides detail for how harmonics relate to consonance, timbre, and pitch. We use the harmonics structure to motivate the design of filters to capture musical features known to impact emotion.

\subparagraph{Pitch Class.}
We organize the harmonics-based filters around the 12 pitch classes of Western music---A, A$\sharp$, B, C, C$\sharp$, D, D$\sharp$, E, F, F$\sharp$, G, G$\sharp$---since instruments are tuned to these pitch classes.\footnote{In Western music, instruments are typically tuned in reference to A4 being 440 Hz to establish typical intervals between instruments.} Table \ref{table:pitch_f0} lists the fundamental frequency of the lowest pitch in each pitch class for nearly all instruments within human hearing, and we use these frequencies to construct the filters.

\begin{table}[htbp]
  \small
  \centering
  \caption{Fundamental Frequency of Lowest Pitch in each Pitch Class}
    \begin{tabular}{lcccccccccccc}
      \hline 
        Pitch & A & A$\sharp$ & B & C & C$\sharp$ & D & D$\sharp$ & E & F & F$\sharp$ & G & G$\sharp$\\
      \hline 
      Frequency (Hz) & 27.50 & 29.14 & 30.87 & 32.70 & 34.65 & 36.71 & 38.89 & 41.20 & 43.65 & 46.25 & 49.00 & 51.91 \\
      \hline
    \end{tabular}
    \label{table:pitch_f0}
\end{table}

\subparagraph{Harmonics Filter Design.}
We now detail the steps to develop the harmonics-based convolution filter for a given pitch class. The unique characteristic of the filter is that it focuses on specific non-contiguous regions of the spectrogram to flexibly learn a wide variety of features. The filter uses ``blinders" to select frequencies input into the CNN, and a convolution size that considers a large range of frequencies at each time frame. The term ``blinders" refers to the matrix operation that selects and weights the mel bands in the mel spectrogram prior to convolution, whereas ``harmonics filter" refers to the combination of the blinders with convolution. Steps (i) - (v) below are done separately for each pitch class.

(i) \textit{Calculate pitch class filter frequencies:} We use the set of frequencies defined in \cref{eq:harmonics} to design blinders that retain only the frequencies of interest. Beginning with the lowest fundamental frequency of pitch class $p$ in Table \ref{table:pitch_f0}, we calculate the set of harmonic frequencies associated with $p$. For example, the lowest A has a fundamental frequency of $f_0=27.50$ Hz, so the A harmonics $\omega_n = n f_0$ are $1 \times 27.50$ = 27.50 Hz, $2 \times 27.50$ = 55.00 Hz, $\ldots$, $801 \times 27.50$ = 22,027.50 Hz. The frequencies span the human hearing range and cover the range of the STFT spectrogram (i.e., up to 22,050 Hz). 

(ii) \textit{Calculate frequency bands:} We calculate frequency bands around each $\omega_n$. The underlying rationale is that the human ear cannot distinguish frequencies within a small band. The exact size of the band depends on a number of factors, including duration, intensity, and frequency, so the width of the bands does not follow a rule. We calculate 1 Hz bands centered around each frequency $\omega_n$.

(iii) \textit{Calculate STFT indicator column:} We identify the match between STFT bins and the frequency bands corresponding to the pitch class. To match the STFT bins with the set of frequency bands, we create an STFT indicator column such that each element is equal to one when the center frequency of the STFT bin falls within one of the frequency bands. STFT bins that are not close to the frequency bands may not be chosen.

(iv) \textit{Construct mel blinders:} We multiply the STFT indicator column by the mel filter bank to generate a mel weight column that has $N_{mel}=256$ dimensions. Repeating the mel weight column over the time dimension generates the mel blinders. Finally, we multiply the mel spectrogram by the mel blinders to generate the input to the CNN's convolution layer. Appendix \ref{section:melblinders} details the steps.

(v) \textit{Implement convolution filter:} There are a few others choices to be made before we can directly apply the convolution filter. For a spectrogram, the convolution filter height determines the number of frequency bins included and the width determines the number of time frames. It is common practice to design the convolution filter to also have a depth (i.e., multiple channels) so that multiple features can be learned simultaneously.\footnote{If height and width represent the $y$- and $x$-axes, channel can be thought about as the $z$-axis. The loss function incentivizes the model to learn different features over different channels. So for example, for images, one channel might learn to detect vertical edges, a second channel horizontal edges, and a third channel texture. The optimal number of channels is determined empirically. The flexibility to learn different features is a large part of why CNNs have been so successful in image classification.} The stride specifies how much the filter slides over the image before performing another operation.\footnote{A small stride captures more fine-grained information but also requires more computational and resource costs.} Since the perception of harmonics is based on the interaction of all frequencies at any given time, we set the filter height to the height of the spectrogram and both the filter width and stride to one time sample.\footnote{Recall that the mel spectrogram has 517 time samples, which is a function of the music clip length, the sampling rate, and the hop length used to construct the spectrogram.}$^{,}$\footnote{In earlier versions of the model, we tried other window lengths and did not find improved predictive performance. We also tried combining filter types but similarly did not see an increase in performance so we chose the most parsimonious model and did not include other filter types.} We learn filters over 8, 16, 32, and 64 channels and find that 32 channels performs the best for prediction, so we set the number of channels to 32. We next apply the $256 \times 1 \times 32$ (number of frequency bins $\times$ number of time samples $\times$ number of channels) convolution filter to the transformed mel spectrogram. The output of convolution on one channel is called a feature map, which is created for each channel.\footnote{This convolution transforms the $256\times517$ mel spectrogram to a $1\times517$ feature map for each channel.} We construct harmonics-based convolution filters for each of the 12 pitch classes. 

\paragraph{\textbf{(S5) Convolutional Neural Network Architecture:}} 

Our CNN to predict music emotion consists of multiple types of layers, including a convolutional layer, pooling layers, and a fully connected layer. Designing a neural network involves several architectural and hyperparameter choices. We use standard architecture choices as appropriate, and describe the reasoning in Appendix \ref{Asection:CNN_arch}. \Cref{figure:architecture} summarizes the overall architecture. The operationalization of the model with harmonics filters is as follows: 
\begin{enumerate}
    \item For each pitch class: 
    \begin{enumerate}
        \item Apply the non-contiguous harmonics filters from Step (S4).
        \item Batch normalize and take ReLU of the feature maps.
        \item Average pool over time frames and apply dropout.
    \end{enumerate}
    \item Concatenate the hidden layers generated by each of the pitch classes.
    \item Max pool over the pitch classes and apply dropout.
    \item Use one fully connected layer and apply softmax to output a probability distribution over the four emotion quadrants.
    \item Output the quadrant Q1, Q2, Q3 or Q4 with max probability as the predicted label.
\end{enumerate}

\begin{figure}
    \begin{center}
    \caption{Model Architecture}
    \includegraphics[width=\textwidth]{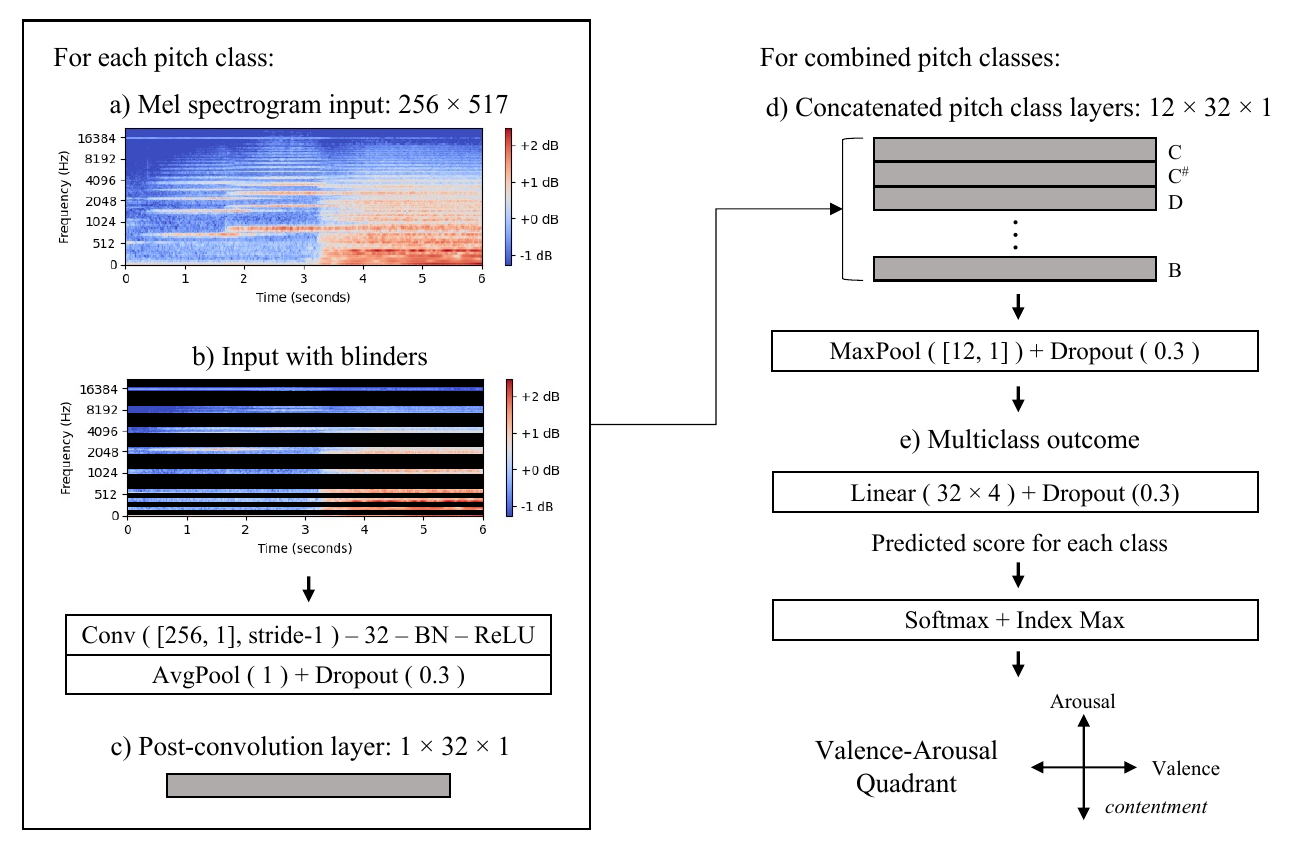}
    \label{figure:architecture}
    \end{center}
\footnotesize Notes: Overview of our proposed CNN architecture with harmonics filters. The input to the CNN is the mel spectrogram. For each pitch class, mel blinders are applied to place structure on what the CNN sees and then convolution and average pooling are applied. The outputs are concatenated together and then max pooled before going through the fully connected layer. The final output is the valence-arousal quadrant prediction.  
\end{figure}

The emotion prediction problem aims to classify an input music clip into one of four valence-arousal quadrants, i.e., a multiclass classification problem. The objective function of the model is to minimize the cross-entropy loss between the predicted and actual outputs. The cross-entropy loss $L_{CE}$ over the set of music clips is:
\begin{equation}
    L_{CE} = -\sum_{i=1}^N \sum_{k=1}^4 y_{ik} log(p(\hat{y}_{ik}))
\end{equation}
\noindent where $k$ represents the quadrant, $y_{ik}$ a binary indicator for whether $k$ is the correct class label for music clip $i$, $p(\hat{y}_{ik})$ the predicted probability that $i$ is of class $k$, and $N$ the total number of music clips. The model learns the weights that minimize the loss.

\paragraph{\textbf{(S6) Predicted Emotion:}}
The model maps each six-second clip into a valence-arousal quadrant. Q1 captures positive valence-high arousal, Q2 captures negative valence-high arousal, Q3 captures negative valence-low arousal, and Q4 captures positive valence-low arousal. To provide a reference emotion for each quadrant, we borrow the notation from \cite{panda2018novel}: Q1---exuberance, Q2---anxiety, Q3---sadness, Q4---contentment. The  predictions over time characterize the dynamics of music emotion.\footnote{Since music is temporal, it is natural to think about using a model that captures dependencies over time. Note, however, that our interest is in classifying emotion for short audio clips, so it is not clear that long-range dependencies will help with predictive performance. An alternative architecture is to incorporate positional information by including an RNN layer in place of average pooling to summarize the feature maps over time. We find that this model results in lower classification performance (F1 = 0.42) than our proposed model without positional encoding. This is consistent with the fact that in \cite{choi2017convolutional} (Figure 3), they also find that RNN marginally reduces the classification accuracy for the two emotions they classify (happy, sad). Due to the short duration of six seconds, there is limited potential to improve performance and the overall performance from adding the RNN deteriorates due to overfitting since the RNN introduces many more parameters. However, including RNN may improve performance with longer duration clips. Note also that in our setting, the emotion of a song may vary over time, so it might not be the case that longer duration clips would always perform better, especially if we are making an assumption that a single clip has only one dominant emotion.}

\subsection{Explainability}\label{section:explainability}

To gain visibility into what the model learns and uses for emotion classification, we use gradient-weighted class activation mapping (Grad-CAM). Grad-CAM uses the gradients of a target class (e.g., Q1) with respect to feature maps, which flow into a given convolutional layer, to produce a heatmap that highlights the regions of the input image that \emph{positively} predict the class \citep{selvaraju2017grad}.\footnote{For example, when classifying an image as a great white shark as opposed to a whale shark, a good image classifier might produce a heatmap that highlights the area with sharp teeth.} Since there is a filter and therefore feature map for each pitch class, we obtain a Grad-CAM heatmap for each pitch class, which are $1\times517$ images (517 representing the number of time samples). The heatmap brightness value for clip $k$ and pitch class $p$ at each time sample $t$ is:
\begin{equation}
    b^j_{kpt} = ReLU\left(\sum_f \alpha^j_{f} A_t^f\right) \text{ where}\quad \alpha^j_f = \frac{1}{Z} \sum_t \frac{\partial y^j}{\partial A_{t}^f}.
\end{equation}
In this expression, $j$ represents the class (quadrant), $A_t^f$ the feature map for channel $f$ at time $t$, $y^j$ the score for class $j$, and $Z$ the feature map size (517).\footnote{This equation is adapted from \cite{selvaraju2017grad} Equations 1 and 2 to our setting where the feature maps for each channel have only one dimension. ReLU takes the max of 0 and the input to the function.} The brightness value is a linear combination of the feature maps $A_t^f$ and their importance $\alpha^j_f$ in predicting class $j$.

We stack the heatmaps associated with each of the 12 pitch classes to obtain an overall heatmap in which the y-axis represents the 12 pitch classes and the x-axis represents time. Color denotes $b^j_{kpt}$, which captures the importance of the non-contiguous frequencies within each pitch class towards the classification into quadrant $j$, and brighter colors capture larger values. Let $B_{k}^{j} = \sum_p \sum_t b_{kpt}^{j}$ represent the sum of the heatmap brightness values over all pitch classes and time samples with respect to quadrant $j$ for clip $k$. Let $\bar{B}_{j} = \frac{\sum_k B_{k}^{j}}{n_j}$ represent the average brightness of all clips with true quadrant label $j$.\footnote{The number of clips with true quadrant label $j$ is ${n_j}$.} 

Combining the structure of the harmonics filters with the Grad-CAM heatmaps enables us to make sense of what the model learns and uses for classification. The harmonics filters retain frequencies relevant for a number of mid-level musical features, including consonance, which is associated with the presence of harmonic combinations of frequencies. Consonance is of particular importance for music emotion and so we expect the filters to not only learn consonance, but also for consonance to play an important role and contribute significantly to emotion classification. Recall that consonance is associated with positive valence and low arousal; while dissonance is associated with negative valence and high arousal \citep{gabrielsson2016relationship}. Since brightness in the heatmaps (i.e., $b^j_{kpt}$) summarizes the positive contribution of all the harmonic frequencies (by pitch class) at a given time to the classification of the target class, we expect high brightness for Q4 and low brightness for Q2. In other words, we expect brightness in the heatmaps to capture consonance. To assess whether the model learns consonance and uses it for classification, we form two sets of hypotheses regarding the brightness of the Grad-CAM heatmaps based on music theory.

First, for a single clip, we expect consonant parts of the clip to have high brightness with respect to Q4 (i.e., $b^4_{kpt}$) since consonance is associated with Q4. On the flip side, consonant parts of the clip should have low brightness with respect to Q2 (i.e., $b^2_{kpt}$) since dissonance is associated with Q2 \citep{gabrielsson2010role}. We expect the brightness with respect to Q1 and Q3 to be between those of Q2 and Q4. Second, across clips in each quadrant, we expect that for a given level of arousal, positive valence clips (based on the true labels) should on average have brighter heatmaps than negative valence clips since consonance is associated with positive valence. Similarly, for a given level of valence, low arousal clips should on average have brighter heatmaps than high arousal clips since consonance is associated with low arousal \citep{gomez2007relationships}. We therefore expect to observe the following pattern: $\bar{B}_{1} > \bar{B}_{2}$, $\bar{B}_{4} > \bar{B}_{3}$, $\bar{B}_{4} > \bar{B}_{1}$, and $\bar{B}_{3} >\bar{B}_{2}$. We assess these hypotheses in \S \ref{section:explainability_empirical}.

Finally, given that consonance would separate out Q4 (high) and Q2 (low), the remaining issue is how the model separates Q1 and Q3. For this, we look at the features learned by the model that are the most important for classification into the quadrants. The learned features correspond to the outputs from the convolutional layer (post-max pooling) that are input to the fully connected layer (the input to Step e in Figure \ref{figure:architecture}). More specifically, we examine whether the learned features most important in predicting Q1 and Q3 correlate to musical features associated with harmonics known to be able to differentiate arousal levels (Appendix \ref{section:harmonics_and_features}). In \S \ref{section:explainability_empirical}, we find the learned features to be connected to rolloff, spectral centroid, and spectral skewness. High arousal is associated with the presence of higher harmonics and higher pitch, which correspond to higher rolloff, higher spectral centroid, and lower spectral skewness. Having identified these features predictive of arousal, we expect learned features positively correlated with rolloff and spectral centroid and negatively correlated with spectral skewness to be positively correlated with the likelihood of a Q1 classification, and negatively correlated with the likelihood of a Q3 classification.

\subsection*{Benchmark Deep Learning Models for Comparison}\label{section:benchmark_models}

We train several benchmark deep learning models for comparison. The models can be characterized as either atheoretical or musically-motivated and focused on low-level features. In contrast to our model, these models only focus on contiguous regions of the spectrogram.

The atheoretical models include CNN with ($n\times n$) square filters, or with rectangular filters, both tall and skinny ($2n\times n$) and short and wide ($n\times 2n$). These are  borrowed from image recognition models as in \cite{krizhevsky2012imagenet}. CNNs with square filters have been fine-tuned to reflect how we see and recognize images but these models do not represent how we hear and process audio. Therefore, square filters are atheoretical from the perspective of acoustic physics. Square filters capture some audio features but it is unclear what these are and how they relate to music emotion. Compared to square filters, rectangular filters of different shapes might allow us to capture features that span a larger portion of the frequency space or a larger portion of the time space.

The musically-motivated models that focus on low-level features include CNN models with filters designed to extract either frequency or time features, first proposed by \cite{pons2016music}. Tall and skinny ($a\times 1$) filters are designed to capture timbral features across the frequency spectrum, e.g., a specific combination of notes, while short and wide ($1\times b$) filters are designed to capture temporal features, e.g., tempo. \cite{pons2016music} apply these ideas to ballroom genre classification and find that, individually, these filters do not perform as well as a CNN with square filters, but that combining the two types of filters with an additional fully connected layer results in comparable performance. We provide additional implementation details for the benchmark models in Appendix \ref{Asection:benchmarks}.

We note that our CNN model with harmonics filters is more parsimonious than the benchmark deep learning models. While our model has 100,000 trainable parameters, the CNNs with square and rectangular filters have nearly 5 million and 10 million, respectively. The CNN with frequency filters has 1.5 million, the CNN with time filters has 2.2 million, and the CNN with frequency-time filters has 3.8 million parameters. With the additional structure, the harmonics filters learn features relevant to emotion recognition as effectively as others models, but with fewer parameters (as we show in \S \ref{section:modelperformance}). As described earlier, this has advantages in generalizability and robustness.

\section{Empirical Analysis}\label{section:training}
We begin by describing the datasets used to train the models. We then report the performance of our proposed architecture with harmonics filters and compare it against benchmark models proposed in the literature. Finally, we show how our model is explainable using gradient-based model visualizations and compare it to visualizations generated by other CNN models which use atheoretical and low-level filters.

\subsection{Datasets}\label{section:datasets}
We combine two public datasets compiled by music emotion researchers, which serve complementary purposes in our analysis: Soundtracks \citep{eerola2011comparison} and the MediaEval Database for Emotional Analysis in Music (DEAM) \citep{aljanaki2017developing}.

The Soundtracks dataset consists of 360 excerpts from movie soundtracks that range in duration from 10 to 30 seconds. One benefit of movie soundtracks is that they are composed to elicit emotion. The music clips are instrumental and do not contain any lyrics, dialogue, or sound effects. The clips were chosen to: 1) elicit either a discrete emotion or to be high or low on valence, arousal, or tension, 2) evoke only a single emotion over the length of the clip, and
3) be unfamiliar to prevent song familiarity from impacting emotion tagging. University students and staff with musical expertise annotated the song emotions, and six annotators tagged each music excerpt.\footnote{Music emotion researchers group emotion into expressed, perceived, and evoked emotion. Expressed emotion refers to the emotion the performer tries to communicate, perceived emotion refers to the emotion a listener perceives from a song, and evoked emotion refers to the emotion a listener actually feels in response to a song \citep{jaquet2014music}. Most often, the emotion of interest is evoked emotion but because of its subjectivity researchers typically build datasets that use perceived emotion labels, as is the case for Soundtracks and DEAM. Perceived and evoked emotion are typically positively related \citep{kallinen2006emotion} so we do not distinguish between the two.} Perceived discrete emotions and valence and arousal were separately annotated on a scale of 1 to 7. Inter-rater consistency (Cronbach's alpha) was 0.92 for valence, 0.90 for arousal, and ranged from 0.66 to 0.93 for the discrete emotions. We split the excerpts into non-overlapping six-second segments. Excerpts selected for the discrete emotions happy, sad, tender, fear, and anger are mapped onto the valence-arousal quadrants such that happy maps to Q1, fear and anger to Q2, sad to Q3, and tender to Q4. The excerpts selected for the dimensional emotions and surprise are converted to the four quadrants by discretizing the valence-arousal space around the midpoint $(4,4)$. All six-second clips from the same excerpt have the same quadrant label since these excerpts were chosen to only evoke a single emotion over the length of the clip. 

The DEAM dataset consists of 1,802 mostly 45-second excerpts of royalty-free music. The music annotations were crowdsourced through MTurk and each excerpt was annotated by at least 10 workers. Perceived valence and arousal were annotated on a scale of $-10$ to $10$ every half second using a graphical interface. The clips were largely unfamiliar to workers. Inter-item consistency (Cronbach's alpha) was 0.28-0.66 for arousal and 0.20-0.51 for valence. In contrast to Soundtracks, the DEAM music was not chosen to elicit a particular emotion, which helps to explain the relatively low annotation consistency.\footnote{The relatively low Cronbach's alpha also highlights the inherent subjectivity of music emotion labeling.} In addition, the DEAM music can vary in emotion over time. DEAM complements Soundtracks by covering more of the valence-arousal space. We include the classical and pop music genres from DEAM since these genres conform to Western music theory. We divide each clip into non-overlapping six-second segments and average the 12 annotations taken every half second to obtain valence-arousal labels. To convert the continuous valence-arousal labels to the four quadrants, we discretize the valence-arousal space around the midpoint $(0,0)$. We thus obtain the same Q1 to Q4 quadrants across both datasets.

Finally, we combine Soundtracks and DEAM. Additional details are provided in Appendix \ref{Asection:data_details}. To improve data balance across the quadrants, we subsample the data so no quadrant has more than 50\% more clips than any other. In total, we have 2,019 six-second music clips distributed 28\%, 19\%, 28\%, 24\% over Q1, Q2, Q3, and Q4, respectively.

\subsection{Model Performance}\label{section:modelperformance}
We use precision, recall, and F1-score, standard measures in the machine learning literature, to evaluate our model. We calculate these metrics for each class (quadrant) and a weighted average by the number of samples in each class determines each overall measure. Table \ref{table:performance_multiclass_main_new} summarizes the performance of the models. The performance metrics are averaged over each fold of the held-out test data from stratified 10-fold cross-validation.\footnote{We create the folds at the song-level rather than the six-second clip-level to prevent data leakage. If clips from the same song are part of both the training and testing data, the training process may pick up some other elements of the song that can be used to predict emotion in the test data, leading to high accuracy, but lower generalizability.} The standard deviations of the performance measures over the ten folds are in parentheses.

\begin{table}[htbp]
  \footnotesize
  \begin{center}
  \caption{Classification Performance - Deep Learning Models}
    \begin{tabular}{lccc}
      \hline \hline
       & Precision & Accuracy/Recall & $F_1$ \\ 
      \hline
      \multicolumn{4}{l}{\textbf{Benchmark Models}} \\
      \multicolumn{4}{l}{\textit{Atheoretical Filters}} \T \\
      Mel - Square   
      & 0.5160 & 0.5003 & 0.4841 \\
      & (0.0568) & (0.0538) & (0.0663) \\
      Mel - Tall Rectangle    
      & 0.5067 & 0.5014 & 0.4809 \\
      & (0.0808) & (0.0590) & (0.0784) \\
      Mel - Wide Rectangle  
      & 0.5232 & 0.4911 & 0.4774 \\
      & (0.0617) & (0.0442) & (0.0585) \\
      \multicolumn{4}{l}{\textit{Theory-based Low-level Filters}} \T \\
      Mel - Time    
      & 0.3597 & 0.3717 & 0.3186 \\
      & (0.1055) & (0.0768) & (0.0890) \\
      Mel - Frequency 
      & 0.5501 & 0.4806 & 0.4703  \\
      & (0.0730) & (0.0532) & (0.0592)  \\
      Mel - Time-Frequency  
      & 0.5058 & 0.4827 & 0.4622  \\
      & (0.0801) & (0.0656) & (0.0771) \\
      \hline
      \multicolumn{4}{l}{\textbf{Proposed Theory-based Mid-level Filters}} \T \\
      Mel - Harmonics 
      & 0.5224 & 0.5049 & 0.5057  \T \\
      & (0.0505) & (0.0478) & (0.0506) \\
    \hline \hline
    \end{tabular}
    \label{table:performance_multiclass_main_new}
    \end{center}
\vspace{2mm}
\footnotesize 
Notes: $\text{Precision} = \frac{\text{True Positive}}{\text{Predicted Positive}}$. $\text{Recall} = \frac{\text{True Positive}}{\text{Actual Positive}}$. $F_1 =  \frac{2\times\text{Precision}\times\text{Recall}}{\text{Precision}+\text{Recall}}$.
Precision is particularly useful when false positives are costly (e.g., spam detection). Recall is particularly useful when false negatives are costly (e.g., disease detection). Accuracy is equivalent to weighted average recall and captures the proportion of correct predictions out of the entire set of data. $F_1$ is useful when we want a balance between precision and recall. The measures are weighted by the proportion in each quadrant.
\end{table} 

In evaluating the performance of the model, it is important to recognize that performance benchmarks for classification using deep learning models can vary significantly by prediction task due to differences in difficulty levels. For example, in a dataset of images of cats and dogs, it might be relatively easy to identify the animal type (cat/dog), but more difficult to identify other labels like age using the same images. Prediction tasks involving \textit{subjective human response} (e.g., emotion recognition, humor detection) typically have lower prediction accuracy than more \textit{objective} recognition tasks (e.g., object identification, instrument identification). An important reason is that for subjective tasks, without objective ground truth, humans have heterogeneous responses, implying less agreement on the ``ground truth" \citep{davani2022dealing}. In such cases, even our best approximation is bounded above by this subjectivity. Music emotion is a challenging classification problem because it is highly subjective. Thus, our model performance should be compared to benchmarks within the class of prediction problems seeking to predict subjective human response. Examples include humor detection for TED talk videos (\cite{chen2017convolutional} obtain an F1 of 0.61 with chance guessing at 0.51) and moral foundation classification (\cite{pavan2023morality} obtain an F1 of 0.45 with chance guessing at 0.19). We note that even though instrument identification and music emotion classification both use music as input data, instrument identification typically has much higher levels of performance.\footnote{For instrument identification, trained ears will have high agreement on which instruments are present, and predicting instruments obtains high F1s, from 0.82 for guitar to 0.99 for drums \citep{blaszke2022musical}. For emotion classification, \cite{liu2017cnn} obtain F1s ranging from 0.41 to 0.60 on the CAL500 and CAL500exp datasets.}

First, we detail the classification performance of the benchmark deep learning models, which use atheoretical and low-level filters. The CNNs with atheoretical square and rectangular filters obtain F1s of around 0.48. While a low-level CNN filter based on time alone does not perform as well as the atheoretical filters, frequency filters perform almost as well as the atheoretical filters. The combination of time and frequency filters performs comparably to the model using only frequency filters, suggesting that frequency-related features play a larger role in eliciting emotion in short music clips relative to temporal features.

Next, we examine the performance of our proposed theory-based mid-level CNN filters that account for harmonics. We find that this model obtains an F1 of 0.51, which is no worse than the best benchmark models, as well as comparable precision and recall measures.\footnote{Appendix \ref{section:confusion} shows the confusion matrices for the square and harmonics filters.}${^,}$\footnote{An alternative input to the mel spectrogram is the STFT spectrogram. Although the literature has most frequently used mel spectrograms since they are more reflective of how humans hear, we also assess the performance of using STFT spectrograms since they are more granular. We find that the average mel model performance measures are slightly higher than those from the STFT spectrograms.}${^,}$\footnote{In Appendix \ref{section:datasize}, we show the effect of dataset size on model performance. We subsample the data and show the change in precision, recall, and F1 for the model with harmonics filters. As expected, we find that more data improves performance but consistent with \cite{brigato2021close}, even very small datasets provide signal to differentiate among classes.}

Overall, the CNN that uses the harmonics filters performs as well as those using atheoretical filters despite having far fewer parameters. The key distinction is that in contrast to the atheoretical filters, our proposed harmonics filters directly connect to theoretical concepts that help explain what the model learns and uses for emotion classification.

\subsection{Model Explainability}\label{section:explainability_empirical}
We now compare the explainability based on Grad-CAM for our CNN model with harmonics filters and other benchmark CNN models.

\subsubsection{Explainability of Harmonics Filters.}

In \S \ref{section:model}, we discussed the patterns to expect in terms of the brightness of Grad-CAM heatmaps based on theory related to harmonics, perception of consonance, and emotional response to consonance. We illustrate explainability around the role of consonance in two ways. First, we compare the heatmaps with respect to all four quadrants for a single clip labeled to be in the high consonance Q4 quadrant. We expect consonant parts of the clip to have high brightness with respect to Q4 and low brightness with respect to Q2. Second, we consider eight distinct out-of-sample clips---two for each of the four quadrants. We expect to see the average ordering of brightness to follow the relative ordering discussed earlier. In addition, we calculate the average brightness of all out-of-sample clips for each quadrant $j$ (i.e., $\bar{B}_{j}$). 

\begin{figure}[h]
    \begin{center}
    \caption{Harmonics Grad-CAM Heatmaps of a Q4---Contentment Clip over Quadrants}
    \includegraphics[width=\textwidth]{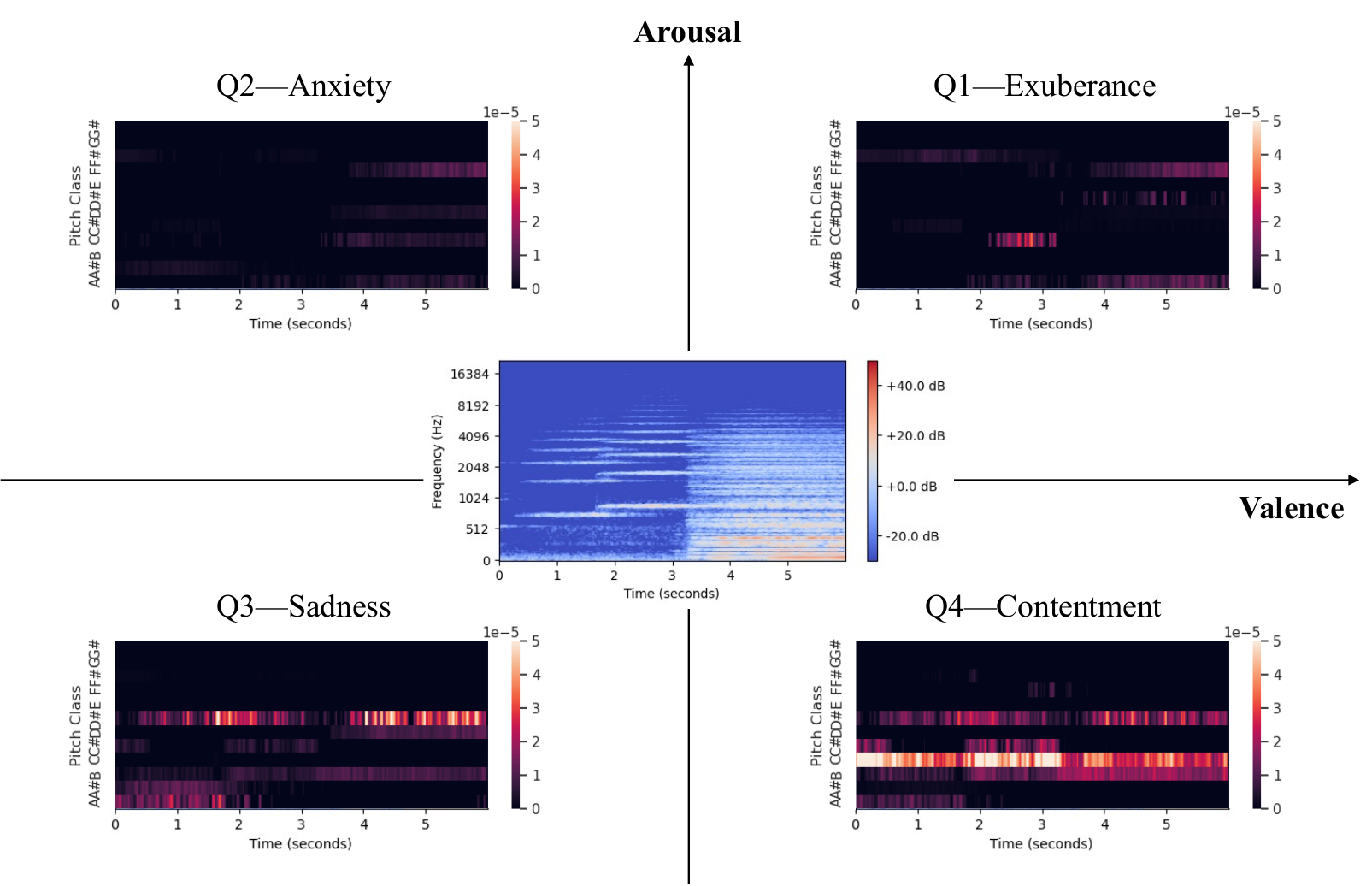}
    \label{figure:soundtracks_183_quadrants}
    \end{center}
    \footnotesize
    Notes: The center image is the mel spectrogram of Soundtracks track 183 seconds 0-6. The images in the four quadrants correspond to the Grad-CAM heatmaps for the classification of the clip into the four emotion quadrants using the CNN model with harmonics filters. The music clip starts off with a violin, producing a harmonic sound that can be seen in the mel spectrogram through the parallel horizontal bars at frequencies that are integer multiples of approximately 740 Hz and 880 Hz. The Q4 heatmap shows that 0.25 to 3.25 seconds most greatly contribute to the Q4 classification, which correspond to periods of overlapping harmonics and consonance. The Q2 heatmap shows this same period does not positively contribute to the Q2 classification. This pattern suggests that brightness in the heatmaps captures consonance.
\end{figure}

First, we look at the heatmaps of a Q4 music clip that we know begins with only frequencies that follow a harmonic pattern, and is therefore consonant. The center image in Figure \ref{figure:soundtracks_183_quadrants} shows the mel spectrogram of a Q4 clip that begins with a violin playing one note, changing notes, and then being joined by additional instruments shortly after three seconds. We know that playing a single violin string produces a harmonic sound, which can be seen in the parallel horizontal bars at frequencies that are integer multiples of approximately 740 Hz between 0.25 and 2.5 seconds (e.g., 740 Hz, 1480 Hz, 2220 Hz), and multiples of 880 Hz between 1.75 and 3.25 seconds (e.g., 880 Hz, 1760 Hz, 2640 Hz). In the Grad-CAM heatmaps, the portions of the clip with harmonic frequencies light up brightly for Q4, indicating these frequencies most greatly contribute to the Q4 classification, but not for Q2. The Q1 and Q3 heatmaps have some points of brightness but not as many as Q4 or as few as Q2. Overall, the Q4 heatmap is much brighter than the other heatmaps. These patterns are consistent with what we would expect to see based on theory if the model learns and uses consonance for emotion classification.

\begin{figure}[h]
    \begin{center}
    \caption{Harmonics Grad-CAM Heatmaps by Quadrant}
    \includegraphics[width=\textwidth]{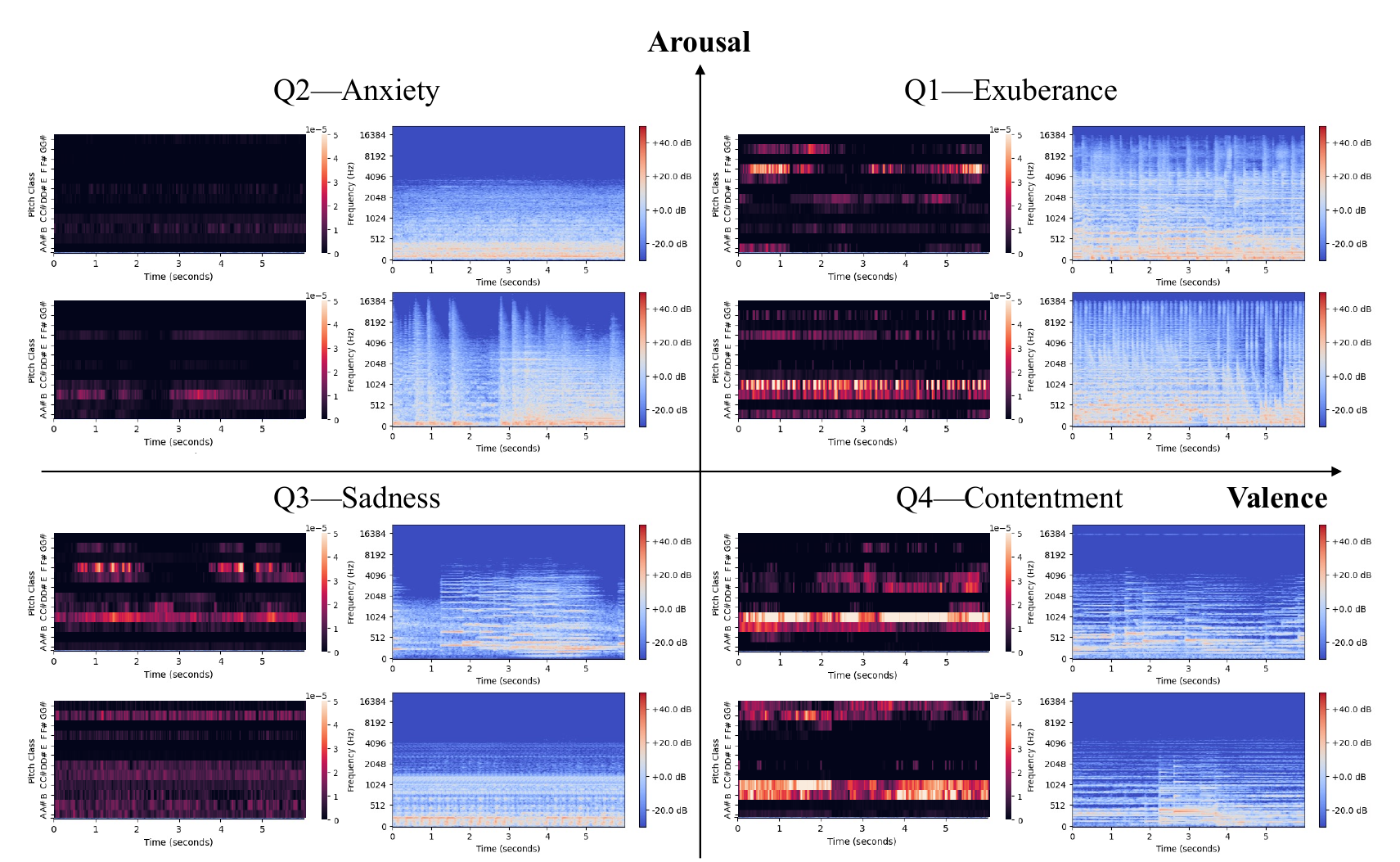}
    \label{figure:grad_cam_mel_quadrants}
    \end{center}
    \footnotesize
    Notes: Within each emotion quadrant, the figures on the right are mel spectrograms of music clips and the figures on the left are their Grad-CAM heatmaps. The following six-second clips are shown: Q1: DEAM song 1811 seconds 27-33 and DEAM song 1892 seconds 21-27; Q2: Soundtracks song 216 seconds 6-12 and Soundtracks song 142 seconds 0-6; Q3: DEAM song 2012 seconds 39-45 and DEAM song 159 seconds 27-33; Q4: Soundtracks song 33 seconds 12-18 and Soundtracks song 48 seconds 6-12.
\end{figure}

Second, we analyze the heatmaps for different music clips for which the model made the same prediction as the true label. Figure \ref{figure:grad_cam_mel_quadrants} shows a few prototypical Grad-CAM heatmaps and their associated mel spectrograms. These heatmaps come from clips in the hold-out sets from ten-fold cross-validation. In general, we observe patterns in line with theory in that positive valence heatmaps are brighter than negative valence heatmaps and low arousal heatmaps are brighter than high arousal heatmaps. 

When we quantify the heatmap brightness across all of the clips for the hold-out sets, we find that the scaled average brightness values based on the true emotion labels are: $\bar{B}_{1}=53$, $\bar{B}_{2}=45$, $\bar{B}_{3}=62$, and $\bar{B}_{4}=72$. The average brightness observed per quadrant is consistent with our hypothesized ordering (i.e., $\bar{B}_{1} > \bar{B}_{2}$, $\bar{B}_{4} > \bar{B}_{3}$, $\bar{B}_{4} > \bar{B}_{1}$, $\bar{B}_{3} >\bar{B}_{2}$), further supporting the consonance-based interpretation of the Grad-CAM heatmaps.\footnote{When we average the brightness values by predicted label, we observe the same ordering. In this case, the scaled average brightness levels are: $\bar{B}_{1}=55$, $\bar{B}_{2}=42$, $\bar{B}_{3}=45$, and $\bar{B}_{4}=90$.}

Explainability builds trust in the model by providing transparency that the model learns patterns consistent with theory, rather than picking up spurious correlations, like discussed in the wolf vs. husky example from \cite{ribeiro2016should} in \S\ref{section:introduction}. The heatmaps highlight areas of consonance, a mid-level feature related to non-contiguous frequencies that is generally not observable by eye, and the patterns of brightness align with the expected relationship between consonance and emotion.\footnote{While we can get a sense of concepts like tempo, loudness, and timbre by reading a mel spectrogram, we cannot visually extract consonance (except in rare cases such as a single violin string being played). Music theorists have proposed several formulas to quantify sensory dissonance based on human hearing and the physics of sound. Our filters enable the deep learning algorithm to learn this relationship as it relates to listener emotion.} 

On the separation between Q1 and Q3, the results are overall consistent with our predictions in \S \ref{section:explainability}. We obtain the features that are the most influential in the classification of a music clip into Q1 and Q3 based on SHAP values.\footnote{SHAP values measure feature importance, capturing the contribution of each feature to the model's output.} The correlations between the two most important learned features for classifying both Q1 and Q3 and musical features known to differentiate arousal are: 0.35 and -0.41 for rolloff, 0.32 and -0.36 for spectral centroid, and -0.32 and 0.44 for spectral skewness. As discussed earlier, the first feature should then be positively correlated with Q1 and negatively correlated with Q3 and the second feature should be negatively correlated with Q1 and positively correlated with Q3 since Q1 is associated with high arousal and Q3 with low arousal. We assess this and as expected, we find that the first learned feature is positively correlated with Q1 (0.45) and negatively correlated with Q3 (-0.47) and that the second learned feature is negatively correlated with Q1 (-0.47) and positively correlated with Q3 (0.51). Thus, being able to separate out high and low arousal helps differentiate between Q1 and Q3.

\subsubsection{Limited Explainability of Atheoretical and Low-Level Filters.}
\begin{figure}
    \begin{center}
    \caption{Square Grad-CAM Heatmaps by Quadrant}
    \includegraphics[width=\textwidth]{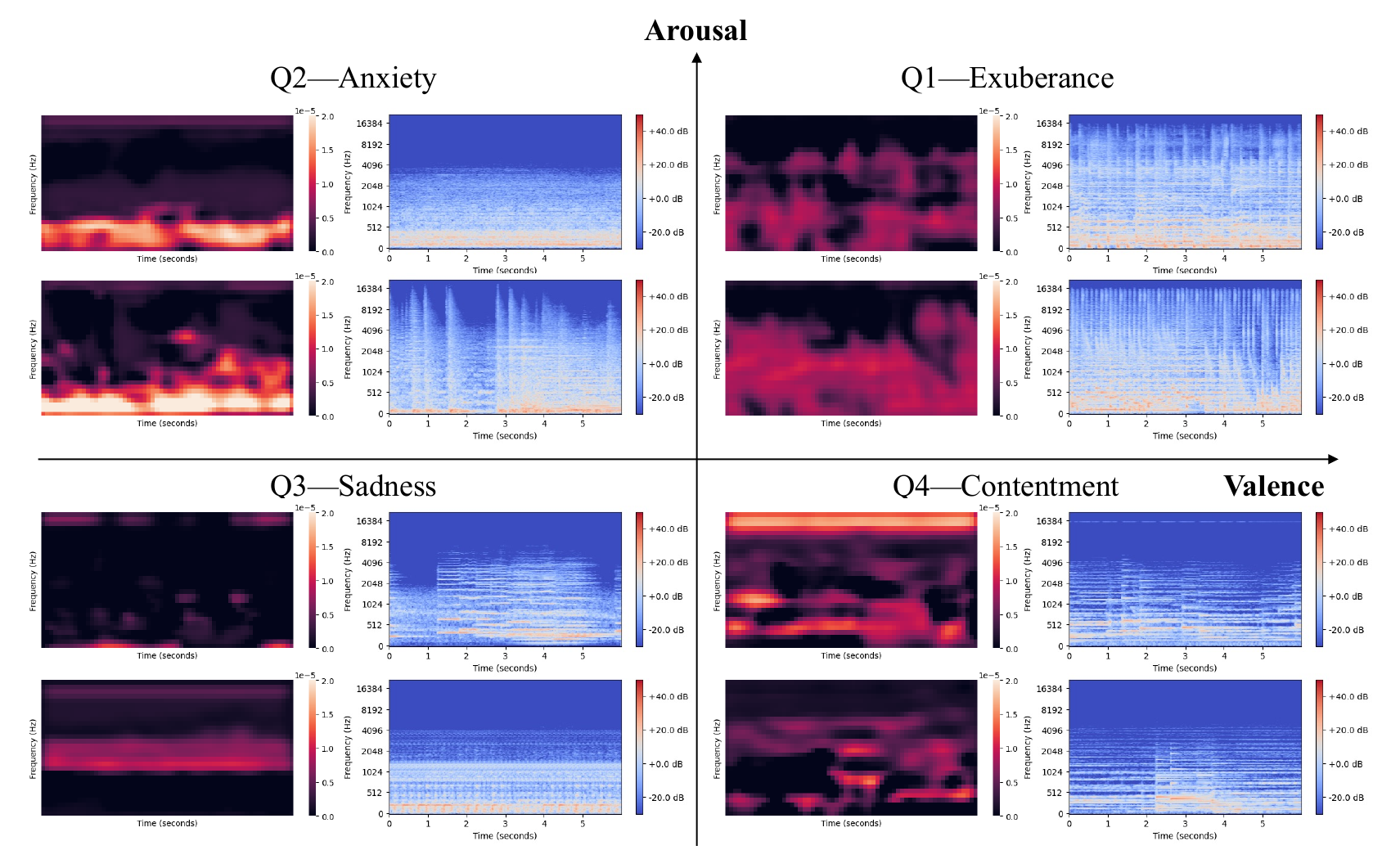}
    \label{figure:grad_cam_mel_square}
    \end{center}
    \footnotesize
    Notes: Within each emotion quadrant, the figures on the right are mel spectrograms of music clips and the figures on the left are their Grad-CAM heatmaps. The clips are the same as those shown in Figure \ref{figure:grad_cam_mel_quadrants}. The bright portions of a heatmap capture the parts of the mel spectrogram that most greatly contribute to the classification of the specified emotion. The heatmap covers the dimensions of the final feature map after convolution. The x-axis captures time and the y-axis captures frequency.
\end{figure}

Grad-CAM visualizations can also be produced for the other filter types. We find that given their low-level focus without specific theory to guide our expectations, it is more difficult to interpret what musical features are captured and connect how they contribute to the classification of a particular class. The Grad-CAM heatmaps for the square filter CNN in \Cref{figure:grad_cam_mel_square} are equivalent to heatmaps produced for an image recognition model. They show which contiguous regions of the input image contribute to the classification of a particular class, providing an idea of the range of frequencies and times contributing to the classification. However, it is unclear what human-understandable musical features are being learned. Note that in contrast to our harmonics filters, pitch classes are not used here.

The heatmaps based on separate time and frequency filters highlight which time periods or frequencies contribute most to a particular classification (see Appendix \Cref{figure:grad_cam_frequency_appendix,figure:grad_cam_time_appendix}). Similar to square filters, it is less clear how to interpret the heatmaps, making it challenging to understand what musical features the models learn. In the models using atheoretical and low-level filters, it is more challenging to determine what musical feature is learned and used for emotion classification. Thus, our contribution in incorporating the non-contiguous theory-based harmonics filters is to provide greater explainability, specifically with consonance, while obtaining similar performance as atheoretical filters.

\subsection{Model Capability to Learn Features}\label{section:otherfeatures}

While our discussion of explainability has primarily focused on a specific and important feature,  i.e., consonance, harmonics filters can also learn other features, as foreshadowed in \S \ref{section:explainability_empirical}. For instance, the combination of harmonic frequencies provides information about pitch perception, timbre (or tone color), and spectral spread and complexity \citep{mcadams2014perception}.

The comparable classification accuracy between the theory-based CNN with harmonics filters and the models with atheoretical filters suggests that the harmonics filters are learning multiple musical features relevant for emotion classification, since consonance alone would not yield such high accuracy. To assess this systematically, we check whether classification performance improves if we add handcrafted features found to predict music emotion. Some handcrafted features, like mel frequency cepstral coefficients (MFCCs), are more atheoretical in that they were not designed for music but were instead designed for speech. Other features, like tempo, are more theory-based in that they were designed with music in mind. Limited improvement in performance would suggest that the model with harmonics filters captures information similar to the handcrafted features. In addition, we assess the correlations between the features learned by the harmonics filters and the handcrafted features. High correlations would suggest that the learned features capture information similar to the handcrafted features.

First, we train two random forest (RF) models using only handcrafted features to show the baseline performance of the features. Random forest is an ensemble learning method known to be both robust and accurate. For music emotion prediction, \cite{unni2022technique} found that random forest provided the best accuracy over a number of machine learning models. The first model uses MFCCs, which have proven successful in music classification, including genre and emotion classification \citep{kim2010music}. We use 13 MFCC coefficients as well as their first and second derivatives, resulting in 39 features. The second model uses 11 handcrafted features highlighted by \cite{panda2018novel} for their ability to predict the four emotion quadrants---we name the set of features the \Panda\ and Appendix Table \ref{table:panda_definitions} summarizes them.\footnote{\cite{panda2018novel} comprehensively consider a wide set of handcrafted features, and identify the ones that most impact classification into the emotion quadrants. Most of the identified features are low-level features that capture tone color/timbre.} 

Next, we combine the handcrafted features with the features learned by our deep learning model that uses harmonics filters. We extract the learned features before the final classification step. We then concatenate these learned features with the handcrafted features, and input this combination to a random forest model to make the final classification. This allows us to test whether the handcrafted features are redundant to the learned features. Another advantage of this approach is that we retain explainability, while being able to add handcrafted features.

\begin{table}[htbp]
  \footnotesize
  \begin{center}
  \caption{Classification Performance - Incorporating Handcrafted Features}
    \begin{tabular}{llccc}
      \hline \hline
      Features & Model & Precision & Accuracy/Recall & $F_1$ \\ 
      \hline
      \multicolumn{5}{l}{\textbf{Benchmark Models with Handcrafted Features}} \T \\
      MFCCs   & RF 
      & 0.4764 & 0.4716 & 0.4615  \\
      && (0.0419) & (0.0446) & (0.0459)\\ 
      \Panda\ & RF 
      & 0.4559 & 0.4563 & 0.4513  \\
      && (0.0637) & (0.0576) & (0.0615)  \\ 
      \hline
      \multicolumn{5}{l}{\textbf{Combined Theory-based Mid-Level Filters + Handcrafted Features}} \T \\
      Mel - Harmonics + MFCCs   & CNN + RF
      & 0.5376 & 0.5278 & 0.5236 \\
      && (0.0525) & (0.0483) & (0.0491) \\
      Mel - Harmonics + \Panda\  & CNN + RF 
      & 0.5361 & 0.5294 & 0.5259  \\
      && (0.0633) & (0.0606) & (0.0605) \\  
      Mel - Harmonics + \TempoF & CNN + RF 
      & 0.5407 & 0.5329 & 0.5295  \\
      && (0.0537) & (0.0493) & (0.0493) \\
    \hline \hline
    \end{tabular}
    \label{table:performance_multiclass_handcrafted}
    \end{center}
\end{table} 

Table \ref{table:performance_multiclass_handcrafted} shows that the inclusion of MFCCs or the \Panda\ marginally improves predictive performance relative to using only the deep learning model (F1 of 0.53 vs. 0.51). This suggests that the harmonics filters do not capture all the information in the handcrafted features but do capture information similar to many of the handcrafted features. When we assess the correlations between the learned features and the handcrafted features, we observe reasonably high values ranging from -0.41 to 0.45 for tone color-related features. This is in line with our expectations since tone color relates to frequency rather than time, and our harmonics filters are specifically designed around frequency. We further find that combining the features learned by the harmonics filters with two time-related handcrafted tempo features improves predictive performance more. 

Overall, the results suggest that our theory-based deep learning model not only learns features that capture similar information to the \Panda\ but also additional features useful for predicting emotion. One advantage of the deep learning model is that each feature does not have to be specifically engineered. However, there is still some value in incorporating handcrafted features related to time.

\section{Application: Emotion-based Ad Insertion in Content Videos}\label{section:application}

Our proposed theory-based deep learning model can be used in a number of real-time \textit{emotion-based} applications by predicting the valence and arousal of music clips. We demonstrate the value with an illustrative application involving emotion-based contextual targeting, where the algorithm determines the optimal emotion-based ad insertion point for a video ad within a content video (e.g., YouTube video) with time-varying emotional content. 

Such emotion-based contextual targeting with automated content matching is gaining importance. Increasing privacy restrictions limit person-specific targeting of advertising, making contextual and content-based targeting for ad placement more relevant and useful.\footnote{Many web browsers have eliminated third-party cookies. In March 2021, Google announced that it would stop tracking the web browsing behavior of individuals (source: \url{https://www.businessinsider.com/google-to-stop-tracking-individuals-web-browsing-precision-ad-targeting-2021-3}).} Further, given the vast amount of UGC available, non-algorithmic approaches to matching ads with content is challenging, if not impossible, to implement at scale. The size of the matching problem is very large, and a platform like YouTube needs to match billions of ads and content videos daily. 

Throughout a content video, emotion often varies over time and so the various ad insertion slots differ in emotion. Since ads also often elicit emotion, we seek to understand how to match the ad to the insertion slot based on emotion. While marketing researchers have considered the overall emotion of content videos for ad matching \citep{coulter1998effects,kamins1991television,puccinelli2015consumers,kapoor2022does}, our focus is on automatically identifying the optimal ad insertion slot within videos that vary in emotion over time.

\textit{Does emotional congruence or contrast work better for ad insertion?} It is an empirical question as to whether ads that are similar to the emotional context of the content video increase or decrease ad attention and memorability. Some behavioral studies have found that emotional congruence is more effective \citep{lee2013interpersonal}, including studies of matching in persuasion and fluency \citep{teeny2021review}. However, other studies have found that consumers have a preference for positive stimuli when feeling negative emotions \citep{andrade2005behavioral} and that perceptual contrast draws attention, suggesting that emotional contrast may be more effective. To answer this empirical question, we conduct a lab experiment in which we exogenously insert ads into content videos at different insertion points. We characterize each ad and each insertion point by their respective emotions, which are based on human tagging. This allows us to create a measure of emotional distance between each ad and each insertion point. We measure ad skip and brand recall and see whether and how emotional distance influences these ad engagement measures. 

Next, after identifying the more effective ad matching strategy, we evaluate whether our proposed theory-based deep learning model can select emotionally appropriate ads based on predicted emotion relative to benchmark models. Given our focus is on music and its effects, we treat the emotion evoked by the background music of the video as a proxy for the emotion evoked by the overall video.\footnote{Film professionals and researchers recognize the importance of music in driving emotion. \cite{nelson2013hollywood} write ``Music plays many roles in film, but it is possible to categorize all of them into two primary functions: creating consonance or dissonance to highlight the film's emotion or narrative."}  We also examine the impact of non-music based emotion in videos. In this setting of automated contextual targeting, the explainability of our model would increase managerial trust in the tool to make reasonable decisions that generalize across a range of different content videos and ads outside of the initial training setting and thus increase confidence in adopting it.

\subsection{Experiment: Is Emotional Congruence or Contrast More Effective?}
We discuss the experimental design and results below.

\subsubsection{Experimental Setup.}
There are multiple ad insertion points, which vary in evoked emotion, within each content video. The outcome variables of interest are ad skip (as a proxy for attention and interest) and brand recall (as a proxy for memorability). We use a full factorial design across four ads, four content videos, and six ad insertion points per content video, yielding a total of 96 experimental cells.

We develop a Qualtrics survey that shows an ad inserted partway through a content video, mimicking the concept of YouTube's mid-roll ads. Similar to YouTube, a ``Skip Ad" button appears six seconds into the ad, allowing participants to skip the remainder of the ad. Upon watching the ad to completion or skipping it, the content video picks up where it left off. Each participant sees only one content video and one ad. After watching the content video, participants are asked questions about the content video and the ad. 

\subsubsection{Content Videos and Ads.}
We select a diverse set of content videos and ads. The content videos are selected to a) contain background music most of the time and b) be long enough to vary in emotion over time and allow for multiple ad insertion points. The videos range from 5.7 to 7.9 minutes in length, include both animated and live-action videos, and include videos with and without speech. Appendix Table \ref{table:content_video_details} provides details on the four content videos. For each content video, we fix six ad insertion points (Time 1, $\ldots$, 6) that are roughly one-minute apart and occur at natural changes in the audio and images, similar to YouTube.\footnote{From YouTube's documentation, ``YouTube’s advanced machine learning technology looks over a large volume of videos and learns to detect the best places for mid-rolls. This is done by evaluating factors like natural visual or audio breaks.'' (Source: \url{https://support.google.com/youtube/answer/6175006?hl=en\#zippy=\%2Cfrequently-asked-questions})} Appendix Table \ref{table:ad_insertion_times} specifies the insertion times.

We select four ads such that each of the four valence-arousal quadrants is represented by the primary emotion in the first six seconds of one of the ads. The emotion of the first six seconds is important, since viewers on YouTube have the option to skip at six seconds. We seek to understand the interaction effect of the content video emotion with the initial ad emotion. The ads include background music, are 30 seconds long, and cover a range of industries.\footnote{For ads originally longer than 30 seconds, we use a 30-second clip to keep the ad length consistent across cells. In these cases, we start and end the ads at natural points.} Appendix Table \ref{table:ad_details} provides details on the four ads.

\textit{Emotion Tagging}: Since we seek to measure emotional distance, we must first characterize the emotion of the content videos and ads. Within the experiment, we use human-tagged emotion as the ground truth. To obtain the emotion tags, we recruit survey participants on Prolific, an online survey platform. We show respondents either the content videos in segments (as defined by Appendix Table \ref{table:ad_insertion_times}) or the first six seconds of the ads and ask them about their valence and arousal levels after watching each clip.\footnote{Each content video received 12-17 tags and each ad received 15-19 tags. Valence and arousal are measured on a scale of 0 to 100 so we convert the valence and arousal levels to valence-arousal quadrants.} With multiple tags per clip, each clip can then be characterized by the distribution of emotion over the quadrants.

\begin{figure}[h]
    \centering
    \caption{Human-Tagged Emotion}
    \begin{subfigure}[b]{0.49\textwidth}
         \centering
         \caption{Content over Time}
         \includegraphics[clip,width=\textwidth]{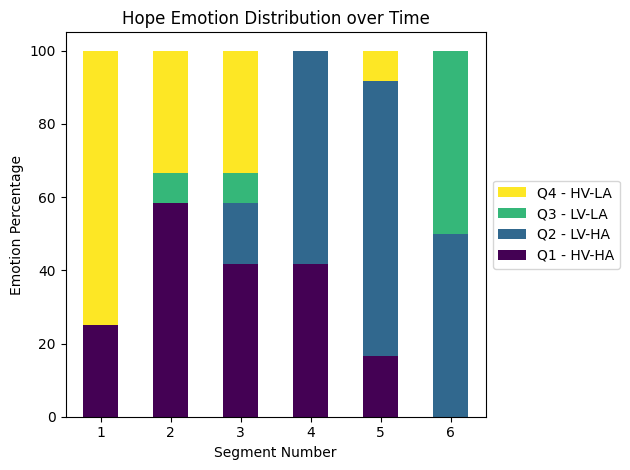}
         \label{figure:hope_emotion}
    \end{subfigure}
    \begin{subfigure}[b]{0.49\textwidth}
         \centering
         \footnotesize
         \caption{Ads}
         \includegraphics[clip,width=\textwidth]{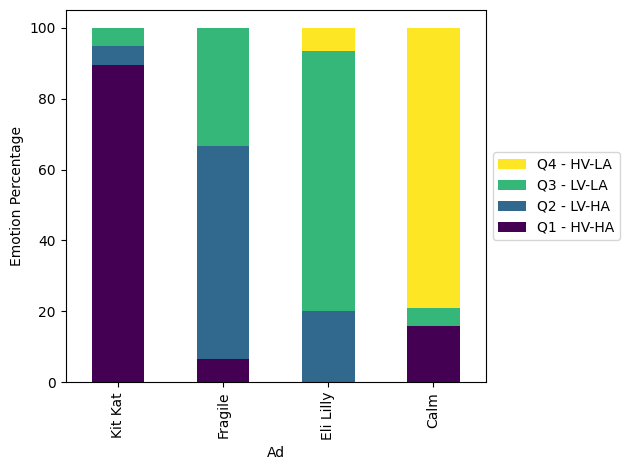}
         \label{figure:ad_emotion}
    \end{subfigure}
\end{figure}

Figure \ref{figure:hope_emotion} shows the emotion distribution over video segments for the content video titled Hope. Viewers largely felt positive (Q1 and Q4) during the first three segments, high arousal (Q1 and Q2) during Segment 4, anxious during Segment 5 (Q2), and negative (Q2 and Q3) during Segment 6, demonstrating large variation in emotion over time. Appendix Figure \ref{figure:content_video_emotions} shows the emotion distributions of all four content videos.

\noindent Figure \ref{figure:ad_emotion} displays the emotion distributions of the first six seconds of each ad. As can be seen, the four ads were selected so that each emotion quadrant would be represented.

\subsubsection{Emotional Distance Measure.}

To measure emotional distance between an ad and an insertion point in the content video, we calculate the Jensen-Shannon (JS) distance. The JS distance is computed between their probability distributions over the emotion quadrants. Let $P_t$ represent the emotion probability distribution of the content video at time $t$, and $Q$ the emotion probability distribution of the ad. JS distance is defined as:
\begin{equation}
    JSD(P_t||Q) = \sqrt{\frac{1}{2}D(P_t||M) + \frac{1}{2}D(Q||M)}
\end{equation}
where $M = \frac{1}{2}(P_t+Q)$ and $D$ is the Kullback–Leibler (KL) divergence. The benefit of JS distance over KL divergence is that it is symmetric between $P_t$ and $Q$ and always finite. The larger the JS distance, the more dissimilar the ad emotion is from the content video emotion at time $t$.

\begin{figure}[h]
    \centering
     \caption{JS Distance between Content Video and Ads}
     \begin{subfigure}[b]{0.48\textwidth}
         \centering
         \includegraphics[clip,width=\textwidth]{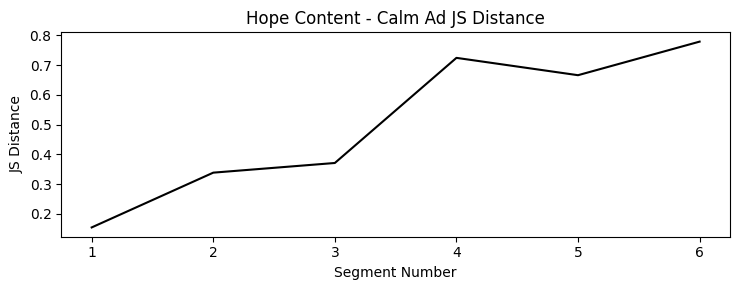}
     \end{subfigure}
     \hfill
     \begin{subfigure}[b]{0.48\textwidth}
         \centering
         \includegraphics[clip,width=\textwidth]{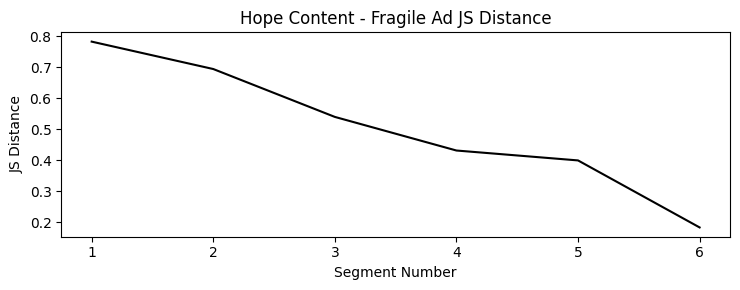}
     \end{subfigure}
    \label{figure:jsd_main}
\end{figure}

Figure \ref{figure:jsd_main} plots the JS distances between the content video Hope and the two ads Calm and Fragile Childhood, across Hope's six insertion points (or segments). For Calm, the minimum distance (i.e., greatest emotional similarity) occurs after Segment 1 and the maximum distance (i.e., greatest emotional contrast) occurs after Segment 6. While for Fragile Childhood, the opposite occurs. Appendix Figure \ref{figure:js_distances} plots the JS distances of the 16 potential combinations of content videos and ads. These plots show that there is large variation in emotional distances between the ads and content videos. The JS distances in the data range from 0.017 to 0.783. 

\subsubsection{Outcomes of Interest.}
Our dependent variables are ``revealed preference''-type metrics and of significant interest to advertisers. Ad skip captures whether someone skips or voluntarily continues to watch the ad. Unaided brand recall captures whether someone paid attention to the ad and its memorability.

An ad skip occurs if the participant presses the ``Skip Ad" button within five seconds of its appearance.\footnote{This is different from YouTube's definition of skip rate (i.e., 1 - view rate). YouTube counts a view as having watched at least 30 seconds of an ad or its duration if it is less than 30 seconds.} This definition captures the emotion interaction of the content video at the time of ad insertion and the first six seconds of the ad. We capture recall in the set of questions asked to participants after watching the content video. In the survey, the brand shows as video information when the ad begins and disappears after three seconds. Therefore, even if participants skip the ad, they are still exposed to the brand. 

\subsubsection{Experimental Results.}
Each participant is randomly assigned to one of the 96 cells.\footnote{Participants were asked to watch a 5-8-minute video and then answer some questions about the video. The survey was limited to Prolific workers who have U.S. citizenship, are fluent in English, have an approval rating greater than 97\%, and have completed at least 50 previous Prolific tasks. Workers who tagged the content video and ad emotions were excluded from participating in the experiment. The survey took on average 11.6 minutes to complete and each participant was paid \$1.80 for their time. Participants that failed the attention checks were excluded.}
Across all content videos, ads, and insertion points, 1,413 participants on average skipped 42.0\% of ads (viewed 58.0\%) and correctly recalled 45.2\% of brands. As expected, the correlation between skip and recall is negative, with a value of $-$0.23 (p-value $<$ 0.01).

To determine the impact of emotional distance on skip and recall, we regress a binary indicator for skipping (or a binary indicator for correctly recalling) on JS distance, controlling for covariates. We estimate the following regression equation:
\begin{equation}
    f(y_{ijt}) = \alpha + \beta JSD_{ijt} + \gamma X_{ijt} + \epsilon_{ijt}
\end{equation}
\noindent where the outcome $y_{ijt}$ represents a skip indicator or a correct recall indicator for ad $i$ at insertion point $t$ within content video $j$, $JSD_{ijt}$ represents the JS distance between the emotion of the content video at point $t$ and the ad, $X_{ijt}$ represents covariates (i.e., ad, content, time of insertion point), and $\epsilon_{ijt}$ the error term. 

For ease of interpretation, we assume a linear probability model for skip and recall and report the least squares coefficients in Table \ref{table:ad_insertion_regression}. The results retain the same signs and levels of statistical significance when we assume a logit model. Table \ref{table:ad_insertion_regression} columns (1) and (4) do not include any covariates. We find that greater emotional distance increases the probability of skipping and decreases the probability of recalling the brand. Columns (2) and (5) include content video fixed effects, ad fixed effects, and a linear time trend for ad insertion time since the times differ across content videos. The results remain robust to including these covariates. Finally, columns (3) and (6) include a second-order polynomial in time and the results do not qualitatively change.\footnote{We also assess the impact of emotional distance on log view time and find that the regression coefficients are negative and statistically significant at the 0.05-level. We focus on skip because we are interested in understanding the interaction of the content video and the beginning of the ad, for which ad skip is a better proxy.} Taken altogether, the JS distance coefficients suggest that emotional congruence between the ad and the insertion point is more effective than emotional contrast for ad engagement.
 
\begin{table}[!htbp]\centering
\def\sym#1{\ifmmode^{#1}\else\(^{#1}\)\fi}
\caption{Effect of Emotional Distance on Ad Engagement}
  \label{table:ad_insertion_regression} 
  \footnotesize
    \begin{tabular}{@{\extracolsep{5pt}}lccc|ccc} 
    \hline \hline 
    Outcome: 
    & \multicolumn{3}{c}{\textit{I(Skip)}} & \multicolumn{3}{c}{\textit{I(Recall)}} \\ 
    & (1) & (2) & (3) & (4) & (5) & (6) \\
    \hline 
    JS Distance & 0.188$^{**}$ & 0.196$^{**}$ & 0.206$^{***}$ 
    & $-$0.179$^{**}$ & $-$0.166$^{**}$ & $-$0.168$^{**}$ \\ 
    & (0.075) & (0.078) & (0.079) &
    (0.077) & (0.077) & (0.077) \\
  Time & & 0.001$^{***}$ & 0.001$^{***}$
  & & $-$0.001 &  $-$0.001 \\ 
  & & (0.001) & (0.001) 
  & & (0.001) & (0.001) \\
  Time$^2$ & & & $-$0.001
  & & & 0.001 \\ 
  & & & (0.001) 
  & & & (0.001)   \\
    \hline 
    Content FE & N & Y & Y & N & Y & Y  \\
    Ad FE & N & Y & Y & N & Y & Y  \\
    R$^2$ & 0.004 & 0.035 & 0.036 
    & 0.003 & 0.092 & 0.092 \\ 
    \hline \hline
    \multicolumn{7}{l}{\textit{Notes:} $^{*}$p$<$0.1; $^{**}$p$<$0.05; $^{***}$p$<$0.01; 1,413 observations for all regressions;} \\ 
    \multicolumn{7}{l}{robust standard errors used}\\
    \end{tabular} 
\end{table} 

\subsection{Ad Insertion Automation}
Our theory-based and benchmark models can predict the emotion distributions of the ads and content to calculate emotional distances at scale. We use these to compare ad skip and recall outcomes from showing each ad at the most emotionally similar ad insertion point based on the predictions of the models.

\subsubsection{Calculating Model-predicted Emotional Distances.}
We transform the first six seconds of audio of each ad and 30 seconds of audio before each ad insertion point in the content videos into mel spectrograms. For the content videos, we break the 30-second clips into five six-second clips. We use the models to predict the emotion distribution of each six-second clip.\footnote{For the deep learning models, the final softmax layer generates a probability distribution over the four valence-arousal quadrants. Instead of selecting the highest probability quadrant, we retain the probability distribution.} For the 30-second content video clips, we average over the five predicted emotion distributions associated with each six-second clip. Using the 24 content emotion distributions (four content videos $\times$ six insertion points) and four ad emotion distributions, we calculate the JS distances between the ads and the content at the six insertion points. For each combination of model, content video, and ad, we determine which ad insertion point is the most emotionally similar.\footnote{For example, we find that for the content video Hope and ad Fragile, the Mel - Harmonics model suggests that the sixth ad insertion point is the most emotionally similar to the ad.} 

\subsubsection{Skip and Recall Rates.}
From the experiment, we have the skip and recall rates for the 96 experimental cells. For each model, we average the skip and recall rates as well as the ground-truth JS distances (based on human-tagging) of the most emotionally similar insertion point for each of the four ads in each of the four content videos as predicted by each model and show the results in  Table \ref{table:automatic_insertion}. Note that these average skip and recall rates are based on a simulation of ad insertion automation using the experimental data. 

\begin{table}[htbp]
  \footnotesize
  \centering
  \caption{Ad Insertion Automation Results}
    \begin{tabular}{llccc}
      \hline \hline
      Feature & Model & JS Distance & Skip & Recall\\
       & &  & Rate & Rate \\
      \hline 
      \multicolumn{4}{l}{\textit{Atheoretic Filters}} \T \\
      Mel - Square & CNN & 0.510 & 42.0\% & 45.7\% \\ 
      Mel - Tall Rectangle & CNN & 0.450 & 43.5\% &  47.3\%  \\ 
      Mel - Wide Rectangle & CNN & 0.419 & 44.5\% & 45.1\% \\ 
      \multicolumn{4}{l}{\textit{Theory-based Low-level Filters}} \T \\
      Mel - Time & CNN & 0.437 & 40.9\% & 46.6\%  \\ 
      Mel - Frequency & CNN & 0.466 & 46.9\% & 45.9\%  \\ 
      Mel - Time-Frequency & CNN & 0.446 & 44.3\% & 48.6\%  \\ 
      \multicolumn{4}{l}{\textit{Proposed Theory-based Mid-level Filters}} \T \\
      Mel - Harmonics & CNN & 0.429 & 42.0\% & 48.9\% \\
      Mel - Harmonics + \TempoF & CNN + RF & 0.395 & 40.7\% & 48.8\% \\ 
      \hline \hline
    \end{tabular}
    \label{table:automatic_insertion}
\end{table}

Our proposed theory-based model using harmonics filters selects insertion points that are relatively high in emotional similarity (i.e., low JS distance) compared to the other deep learning models. These ad insertion points generate relatively favorable skip and recall rates, suggesting that our model can be useful in automated emotion-based ad insertion. The models that includes tempo features, the best performing model from \S \ref{section:otherfeatures}, further improves upon these results.

\subsubsection{Incorporating Other Video Modalities.}\label{section:other_modalities}

Our primary analysis has focused on using emotion evoked from music. However, with videos, emotional content may be present across multiple modalities (e.g., facial expressions, text of speech). Multimodal emotional content can also be used to predict emotional distance. When the videos have human faces, we can use publicly available tools to estimate emotion from facial expressions. Similarly, emotional content can also be obtained from voice tonality and speech text. 

In our application, we observe that speech or human faces might not be present in each content video or in the first six seconds of ads, implying using speech or facial emotion is not always feasible. For the videos with human faces, we assess the skip and recall rates when including face emotion alongside music emotion. We could not include speech emotion since there was not enough speech in the first six seconds of the ads. We find mixed evidence of the value of including face emotion. Including face emotion slightly improves the recall rate but hurts the skip rate. Appendix \ref{section:ms_azure} details the analysis.

Overall, we find that there is potential in incorporating emotion information from images and text, but the existing tools are limited in their ability to extract emotion information from short clips (i.e., first six seconds of ads) and animated videos. However, this is a moving target and as these methods steadily improve, these findings could well change.

\subsection{Managerial Implications}

Past studies have provided evidence that emotional ads impact attention and memory \citep{cohen2018nature,holbrook1987assessing}. The results of this study support the theory that emotional similarity decreases ad skipping and increases brand recall. Our method can be used to determine time-varying emotion based on the background music of videos, and we show that it performs as well as atheoretical CNN models while being explainable.

We demonstrate the value of our model in a video advertising setting by mapping music to emotion to determine the optimal emotion-based insertion point within a content video. In practice, we expect emotion to be used as a complement to other ad targeting variables. 

Our proposed model could also be useful in a number of other applications. For example, existing Spotify playlists built around a unifying emotion are based on the overall emotion of a song. However, one quarter of songs are skipped in the first five seconds, so the interaction of the ending of one song and the beginning of the next is a critical point for a listener's decision to continue with a playlist.\footnote{https://www.theguardian.com/music/2014/may/07/one-quarter-of-spotify-tracks-are-skipped-in-first-five-seconds-study-reveals} Our model can quantify the emotional match between the end of one song and the beginning of the next to allow for continuity (or contrast) in the listener's emotional experience. The classifier can also be used in contexts that match music with other forms of unstructured data. For example, a text classifier can be used for a news article while our model can be used for the video ad. More broadly, any setting that involves emotion and requires music choice (e.g., call waiting music) could benefit from a music emotion classifier.

\section{Conclusion}\label{section:conclusion}
Our research contributes to the literature that studies consumer response to music, which represents one type of unstructured data. Music is pervasive in customer interactions with firms. From music in ads to hold music for call centers, from workout playlists to background music in retail stores, customers engage with music in a variety of ways. The exponential growth of user-generated content has created a huge quantity of high-dimensional data, and automated prediction of music-evoked emotion at scale can be helpful for a variety of marketing decisions. However, unlike other unstructured data, such as text (e.g., \citealt{toubia2021quantifying,wang2021attribute}), images (e.g., \citealt{liu2020visual,dew2022letting,zhang2022can,troncoso2022look,huang2022variety,sisodia2023automatic}), and video (e.g., \citealt{yang2021first,chakraborty2022ai}), music has received relatively little attention in the marketing literature. 

We develop a CNN to classify the emotion evoked by music. Our framework integrates a number of theoretically motivated elements to develop harmonics-based filters, combining the physics of music, human hearing of sound, as well as human perception of music. Our approach achieves similar classification performance to that of atheoretical models. In terms of explainability, we exploit specific elements of music theory to construct filters capable of capturing consonance. We visualize the model's prediction process using Grad-CAM, which provides a visual representation of the areas in an image (spectrogram for sound) that contribute the most to the classification into a particular target class. While this provides a degree of transparency, we note that making deep learning models more explainable is an active area of research in machine learning. Finally, we use our music emotion classifier in an application where we match the time-varying emotion in a content video with the emotion of the first six seconds of an ad and show that it performs as well as benchmark models while being more explainable.

We conclude with a discussion of some limitations and suggestions for future research. First, we focus on relatively short music clips in our data and model. This choice is motivated by the application of ad insertion, where typically a video ad is played and the user is allowed the choice to skip six seconds into the ad. While the method in principle applies to clips of any duration, in practice we might consider altering the architecture of the deep learning model to include temporal dependencies. Second, incorporating listener heterogeneity based on demographics could further improve the model's predictive accuracy, but given privacy concerns, we have opted to leave out listener data. Third, investigating the complementarity of unstructured high-dimensional data across modalities would be valuable. In sum, we believe that the growing variety of data in conjunction with explainable models offers rich research opportunities to quantify emotion and its impact.

\section*{Funding and Competing Interests}
All authors certify that they have no affiliations with or involvement in any organization or entity with any financial interest or non-financial interest in the subject matter or materials discussed in this manuscript. The authors have no outside funding to report.

\bibliographystyle{informs2014} 
\bibliography{output.bbl} 

\setcounter{table}{0}
\renewcommand{\thetable}{\Alph{section}\arabic{table}}
\setcounter{figure}{0}
\renewcommand{\thefigure}{\Alph{section}\arabic{figure}}

\begin{APPENDIX}{}

\section{Music Concepts}\label{Asection:definition}

\begin{table}[htbp]
  \footnotesize
  \centering
  \caption{Definitions of Music Concepts}
    \begin{tabular}{p{1.7cm} p{1.5cm}p{12.4cm}}
      \hline \hline
        Concept & Type & Definition \\ 
      \hline 
      Frequency & Physical & The number of cycles a sine wave completes in a second, measured in Hertz (Hz) \\
      Fundamental Frequency & Physical & Lowest natural frequency of a sine wave \\
      Partial & Physical & Any of the sine waves that comprise sound \\
      Harmonic & Physical & A frequency that is an integer multiple of the fundamental frequency \\ 
      Spectrum & Physical & The range of frequencies contained in a signal \\ 
      Musical Interval & Physical & Spacing between two sounds in frequency \\
      \hline
      Pitch & Perceptual & The attribute of sound that allows it to be ordered on a scale from low to high \\
      Note/Tone & Perceptual & A pitched sound \\
      Pitch Class & Perceptual & Set of all pitches that are an integer number of octaves apart \\
      Harmony & Perceptual & Set of pitches played simultaneously \\
      Tonalness & Perceptual & Music that has a specific note on which it is the most stable and at rest \\
      Consonance & Perceptual & A combination of notes that sound pleasant when played simultaneously \\ 
      Dissonance & Perceptual & A combination of notes that sound harsh or jarring when played simultaneously \\ 
      Loudness & Perceptual & The intensive attribute of an auditory sensation, in terms of which sounds may be ordered on a scale extending from soft to loud
      \\
      Timbre & Perceptual & The attribute of auditory sensation in terms of which a listener can judge that two sounds similarly presented and having the same loudness and pitch are dissimilar 
      \\
      \hline \hline
    \end{tabular}
    \label{table:music_definitions}
\end{table}

\section{Music Feature Interpretability}
\setcounter{table}{0}
\setcounter{figure}{0}
\cite{fu2010survey} use top-level labels to describe music constructs humans understand, like emotion and genre. To predict these labels, researchers use audio features, which can be divided by level of music understanding (i.e., interpretability). Low-level music features, obtained directly from the audio using simple mathematical transformations, are not closely connected to musical properties perceived by human listeners. More complex mid-level music features are usually based on transformations of the low-level features and are more closely connected to musical properties perceived by humans.
\begin{table}[h]
  \footnotesize
  \centering
  \caption{Music Features by Interpretability Level \citep{fu2010survey}}
    \begin{tabular}{lll}
      \hline \hline
      Feature Type & Musical Construct & Examples  \\ 
      \hline 
      Top-level labels &  Emotion & Valence, arousal \\
       &  Genre & Pop, rock, jazz \\
       &  Instrument & Piano, violin, flute \\
       \hline
      Mid-level features &  Harmony & Chord sequences \\
       &  Rhythm & Beat histogram \\
       &  Pitch & Pitch histogram, chroma \\
      Low-level features &  Frequency & MFCC, zero crossing rate  \\
       &  Time &  Amplitude modulation, statistical moments \\
      \hline \hline
    \end{tabular}
    \label{table:music_feature_interpretability}
\end{table}
\FloatBarrier

\newpage
\section{Harmonics and Mid-level Musical Features}\label{section:harmonics_and_features}
\setcounter{table}{0}
\setcounter{figure}{0}
We explain how harmonics impact mid-level musical features that influence emotional response to music.

\subsection{Harmonics and Consonance}
Harmony captures the perception of simultaneous pitches and is characterized as being consonant or dissonant. In general, consonant sounds, such as the octave, are considered ``pleasant or restful," while dissonant sounds are considered jarring \citep{sethares2005tuning}.\footnote{A classic example of a dissonant sound is the tritone. The tritone has been used in contemporary movies and music to provide a negative connotation or of something foreboding or fear-inducing \citep{lerner2009music}.} Studies have revealed that consonance and dissonance are not binary categories, but rather opposite ends of a continuum.

\cite{plomp1965tonal} show that unison (two notes with identical frequencies) is the point of global maximum of consonance and specific other two-note frequency intervals form local maxima. Consonance is associated with small integer ratios of pitch frequencies. Music theorists have suggested that the physics underlying consonance is the occurrence of overlapping harmonics \citep{sethares2005tuning}, which occurs with small integer ratios. When a given sound has many overlapping harmonics, it is perceived as being consonant. 

Consonance and dissonance are known to influence emotional response to music. Consonance is associated with positive valence emotion (e.g., tenderness) while dissonance is associated with negative valence emotion (e.g., fear). Consonance is also associated with low arousal emotion (e.g., contentment) while dissonance is associated with high arousal emotion (e.g., fear) \citep{gabrielsson2016relationship}.

\subsection{Harmonics and Timbre}

Timbre is the quality of a sound that distinguishes it from another sound with the same loudness and pitch. Different musical instruments and voices create different patterns of harmonics varying in arrangement and strength when generating the same fundamental frequency, which create their timbre, or the ``acoustic fingerprint" of an instrument or voice \citep{nelson2013hollywood}. Tones with a strong emphasis on higher harmonics (i.e., high multiples of the fundamental frequency) are associated with high arousal while tones with suppressed higher harmonics are associated with low arousal \citep{gabrielsson2016relationship}. Given that the filters are designed around harmonic frequencies, they are capable of learning different timbre patterns that help predict emotion.

\subsection{Harmonics and Pitch}

Pitch is the quality of a sound that allows it to be ordered on a scale from low to high. The fundamental frequency determines the perceived pitch, but the presence and alignment of harmonics affect how clearly the pitch is perceived. According to \cite{gabrielsson2016relationship}, ``High pitch may be associated with expressions as happy, graceful, serene, dreamy, exciting, surprise, potency, anger, fear, and activity. Low pitch may suggest sadness, dignity/solemnity, vigor, excitement, boredom, and pleasantness." The filters are capable of learning different pitch patterns predictive of emotion.

\section{Benchmark Models for Comparison}\label{Asection:benchmarks}
\setcounter{table}{0}
\setcounter{figure}{0}

\noindent \textit{CNN with Square Filters.}
Image recognition CNN models typically use square filters that capture associations across two orthogonal spatial dimensions. Although mel spectrograms visualize music, the vertical and horizontal dimensions represent frequency and time rather than spatial dimensions. Thus, in music, the dimensions have very different meanings and resulting properties.

To operationalize the CNN with square filters, we borrow the architecture based on the VGG image classification model used by \cite{chowdhury2019towards} to classify emotion on the Soundtracks dataset. The model includes nine convolutional layers that primarily use $3\times3$ square filters alongside batch normalization, ReLU, and dropout. After the ninth convolutional layer, the model uses average pooling to summarize the information over different channels and then a fully connected layer to predict valence and arousal. 

\noindent \textit{CNN with Rectangular Filters.}
It is possible that square filters might not be ideal to capture the features of audio data. Thus, we generalize this by using rectangular filters that are relatively more: (a) tall and narrow (specifically, replace each $k \times k$ convolution filter by a $2k \times k$ filter), and (b) short and wide (specifically, replace each $k \times k$ convolution filter by a $k \times 2k$ filter). We make these replacements for both the $5 \times 5$ and $3 \times 3$ filters that are used in our baseline square filter model.

\noindent \textit{CNN with Time and Frequency Filters.}
We compare our proposed mid-level harmonics filters against low-level time and frequency filters proposed by \cite{pons2016music}. We design a model that uses frequency filters, a model that uses time filters, and a model that combines the two types of filters. Frequency filters, which are tall and skinny, are designed to capture timbral features across the frequency spectrum, e.g., a specific combination of notes, while time filters, which are short and wide, are designed to capture temporal features, e.g., tempo. We allow the models additional flexibility by including an additional fully connected layer after pooling and before the final classification. 

\section{Consonance Blinders Transformation}\label{section:melblinders}
\setcounter{table}{0}
\setcounter{figure}{0}
We describe the process to transform the STFT blinders to mel blinders. To simplify the problem, let us assume that the frequency dimension of the STFT is continuous for now. For fundamental frequency $f_0$ we retain the frequencies $f_0$, $f_1=2f_0$, $f_2=3f_0$,..., $f_n=(n+1)f_0$. To account for human auditory perception, we allow for a band of frequencies centered around each frequency. The bandwidths are based on a constant bandwidth so our retained frequencies with bandwidth $\delta$ are of the form: $[f_n - \delta, f_n + \delta]$. 

We allocate the power associated with the frequencies to the mel bands. The mel filter bank maps frequencies to (a maximum of 2) mel bands. \Cref{figure:mel_filter} shows the mapping of frequencies to 20 mel bands (in the paper we use 256 but it is more challenging to visualize). The top of each triangle represents the center of each band. Each triangle represents the weight each frequency contributes to a particular band. For example, the right most triangle maps frequencies ranging from roughly 5,400 - 7,000 Hz to the 20th mel band. Frequencies below 5,400 Hz receive zero weight. The triangles grow wider with higher frequencies because human hearing resolution is worse at higher frequencies.
\begin{figure}[h]
    \centering
    \caption{Mel Filter Bank}
    \includegraphics[width=.4\textwidth]{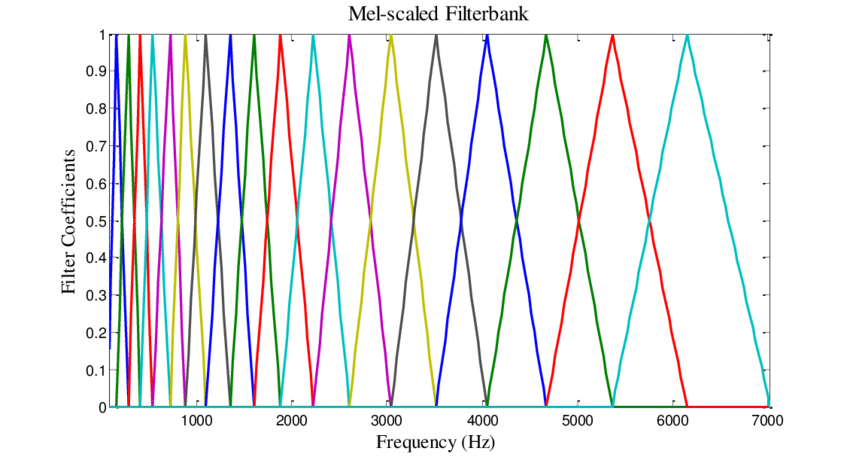}
    \label{figure:mel_filter}
\end{figure}
Let $\beta_j$ represent the function that maps frequencies to mel band $j$ (i.e., the triangles). Let $b^-$ and $b^+$ represent the lowest and highest frequencies, respectively, which map to mel band $j$ and $b$ the midpoint of the two numbers $(\frac{b^-+b^+}{2})$. $\beta_j$ is defined as:
\begin{equation}
    \beta_j (x) = 
    \begin{cases}
      0 & \text{if $x < b^-$} \\
      \frac{x - b^-}{b - b^-} & \text{if $b^- \leq x \leq b$} \\
      \frac{b^+ - x}{b^+ - b} & \text{if $b \leq x \leq b^+$} \\
      0 & \text{if $x > b^+$}
    \end{cases}
\end{equation}
Then the contribution of frequency band $[f_n - \delta, f_n + \delta]$ to mel band $j$, $M_{jn}$, is:
\begin{equation}
    M_{jn} = \int_{f_n - \delta}^{f_n + \delta} \beta_j(x) P(x) \phi(x) dx
\end{equation}
\noindent where $x$ represents frequency, $P(x)$ the power of $x$, and $\phi(x)$ the distribution over frequencies. We assume $\phi(x) \sim U[f_n - \delta,f_n + \delta]$. Since multiple frequency bands could contribute to a single mel band, we sum the contributions so the final power of mel band $j$ is $M_j = \sum_{n} M_{jn}$. The set of power over all $j$ comprises the mel blinders. They highlight which mel bands are input to the CNN and the weight of each band.

\section{Mel - Harmonics CNN Architecture Choices}\label{Asection:CNN_arch}
\setcounter{table}{0}
\setcounter{figure}{0}

We describe a few relatively standard CNN modeling choices and their operationalization in the model.

\begin{table}[h]
  \footnotesize
  \centering
  \caption{Standard CNN Model Elements}
    \begin{tabular}{p{2.3cm}p{13.5cm}}
      \hline \hline
      Element & Purpose  \\ 
      \hline 
      Pooling & Pooling applies a function over all units within a specified filter shape. We evaluate average pooling over time and both average and max pooling over pitch class. Max pooling over pitch class results in a higher F1 than average pooling so we use average pooling over time and max pooling over pitch class in our main model specification. \\
      Batch Normalization & Batch normalization standardizes the inputs (mean zero, standard deviation one) in a mini-batch (data seen each time model parameters are updated). This procedure standardizes the inputs and helps achieve faster training, reduces overfitting, and stabilizes learning. \\
      ReLU & Rectified Linear Activation Unit (ReLU) transforms the output of convolution to allow the model to learn nonlinear relationships. \\
      Dropout &  A form of regularization, dropout randomly removes neurons in specified layers of the neural network each mini-batch based on the specified dropout rate to prevent overfitting and obtain a more robust model. \\
      Fully Connected Layer & Layer that connects every neuron in the hidden layer previous to every neuron in the next layer. \\
      \hline \hline
    \end{tabular}
    \label{table:cnn_model_ingredients}
\end{table}

\section{Dataset Merging Details}\label{Asection:data_details}
\setcounter{table}{0}
\setcounter{figure}{0}

Our dataset is made up of the Soundtracks data combined with the DEAM classical and pop data. Soundtracks and DEAM were annotated on different numeric scales and in this section, we describe how we combine the two. For the Soundtracks data, perceived valence and arousal were annotated on a set of discrete emotions, as well as on bipolar scales (using adjectives), which the researchers transformed to a scale ranging from 1 to 7. The valence extremes were captured by the adjectives pleasant-unpleasant, good-bad, and positive-negative. The arousal extremes were captured by the adjectives awake-sleepy, wakeful-tired, and alert-drowsy. The midpoint of the numeric scale for Soundtracks is (4,4) for valence and arousal. 

We observe that for the excerpts chosen to capture discrete emotions, the labels have high inter-rater consistency for all emotions except for surprise. We therefore use the discrete emotion labels to map these excerpts to the emotion quadrants. For the excerpts chosen to capture dimensional emotions and surprise, we use the valence and arousal labels to map these excerpts to the four quadrants.

For the DEAM data, perceived valence and arousal were annotated continuously on a scale of -10 to +10. Like Soundtracks, the DEAM creators also provided adjectives to describe valence and arousal. The valence extremes were extremely negative/unpleasant to extremely positive/pleasant with neutral in the middle. The arousal extremes were low arousal/calm to high arousal/activated/excited. \cite{aljanaki2017developing} transform the data to range from -1 to +1. The midpoint of the numeric scale is therefore (0,0).

It is important that the Q1 to Q4 emotion labels are based on the scale used for each dataset. For example, Q1 captures positive valence-high arousal emotion. For Soundtracks, this maps to valence $\geq$ 4 and arousal $\geq$ 4 or discrete emotion $=$ happy while for DEAM this maps to valence $\geq$ 0 and arousal $\geq$ 0. It would be incorrect to label Soundtracks using the DEAM scale and vice versa. Essentially, we have standardized the data so that the emotion quadrants mean the same thing for the DEAM and Soundtracks datasets.

\section{Impact of Dataset Size}\label{section:datasize}
\setcounter{table}{0}
\setcounter{figure}{0}

\begin{table}[htbp]
  \footnotesize
  \begin{center}
  \caption{Performance by Dataset Size for Mel - Harmonics Model}
    \begin{tabular}{lccc}
      \hline \hline
       & Precision & Accuracy/Recall & $F_1$ \\ 
      \hline
      25\% of data 
      & 0.5069 & 0.4713 & 0.4571 \\ 
      & (0.0893) & (0.0711) & (0.0598) \\
      50\% of data 
      & 0.5129 & 0.5023 & 0.4976 \\ 
      & (0.0606) & (0.0570) & (0.0601) \\
      100\% of data 
      & 0.5224 & 0.5049 & 0.5057  \\
      & (0.0505) & (0.0478) & (0.0506) \\
    \hline \hline
    \end{tabular}
    \label{table:performance_size}
    \end{center}
\end{table} 

\section{Handcrafted Features}\label{section:handcrafted}
\setcounter{table}{0}
\setcounter{figure}{0}

We use 11 handcrafted features highlighted by \cite{panda2018novel} for predicting the emotion quadrants. 

\begin{table}[htbp]
  \footnotesize
  \begin{center}
  \caption{Top Music Emotion Base Features from \cite{panda2018novel}}
    \begin{tabular}{p{4cm}p{2.75cm}p{1.75cm} p{6.5cm}}
      \hline \hline
        Feature in \cite{panda2018novel} & Feature in MIR Toolbox & Musical Concept & Definition \\ 
      \hline 
      FFT Spectrum - Spectral 2nd Moment (median) & Spectral Spread (median) & Tone Color & Standard deviation of the spectrum; a measure of the spread of the distribution \\ 
      FFT Spectrum - Average Power Spectrum (median) & Spectral Centroid (median) & Tone Color & The geometric center of the spectrum distribution can be an indicator of the ``brightness" or ``sharpness" of the sound \\
      FFT Spectrum - Skewness (median) & Spectral Skewness (median) & Tone Color & The third moment of the spectrum; a measure of the symmetry of the distribution \\ 
      Spectral Skewness (std) & Spectral Skewness (std) & Tone Color & See above \\
      Spectral Skewness (max) & Spectral Skewness (max) & Tone Color & See above \\
      MFCC1 (mean) & MFCC1 (mean) & Tone Color & MFCC offers a description of the spectral shape of the sound \\
      MFCC1 (std) & MFCC1 (std) & Tone Color &  See above \\
      Roughness (std) & Roughness (std) & Tone Color & An estimation of the sensory dissonance \\
      Rolloff (mean) & Rolloff (mean) & Tone Color & Fraction of energy below specific frequency \\
      Spectral Entropy (std) & Spectral Entropy (std) & Tone Color & Shannon entropy offers a general description of the spectral power distribution \\
      Fluctuation (std) & Fluctuation (std) & Rhythm & Estimates the rhythm content based on spectrogram transformed by auditory modelling \\
      \hline \hline
    \end{tabular}
    \label{table:panda_definitions}
    \end{center}
\end{table}

\newpage
\section{Confusion Matrices}\label{section:confusion}
The confusion matrices for the other models are available upon request from the authors.
\setcounter{table}{0}
\setcounter{figure}{0}
\begin{figure}[h]
\caption{Confusion Matrices}
\begin{minipage}{0.45\textwidth}
    \begin{center}
    \subcaption{\small Mel - Square Filters}
    \includegraphics[clip,width=\textwidth]{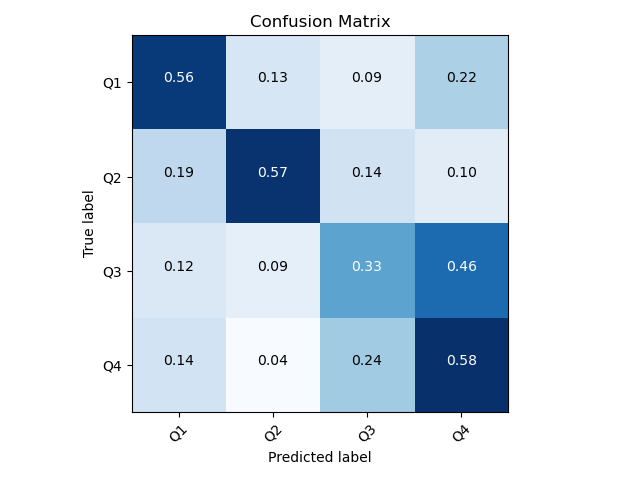}
    \label{figure:rec_wide_confusion_matrix}
    \end{center}
\end{minipage}
\begin{minipage}{0.45\textwidth}
    \begin{center}
    \subcaption{\small Mel - Harmonics Filters}
    \includegraphics[clip,width=\textwidth]{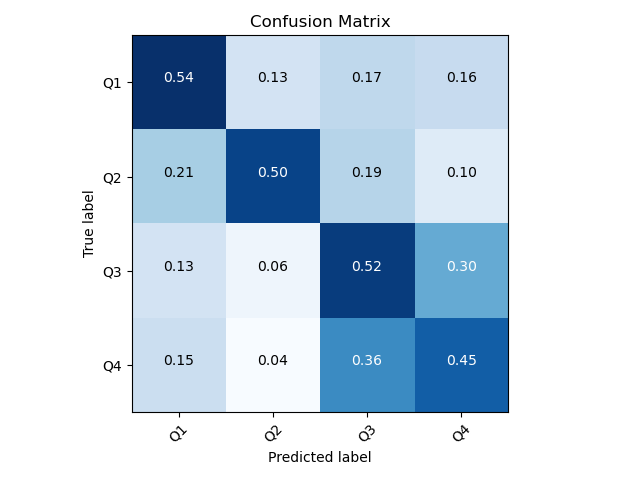}
    \label{figure:harmonics_confusion_matrix}
    \end{center}
\end{minipage}
\end{figure}

\newpage
\section{Grad-CAM Visualizations}\label{section:other_grad}
\setcounter{table}{0}
\setcounter{figure}{0}
\begin{figure}[!h]
    \centering
    \caption{Frequency and Time Grad-CAM Heatmaps}
    \label{fig:enter-label}
         \begin{subfigure}[b]{\textwidth}
             \begin{center}
        \caption{Frequency Heatmaps}    \label{figure:grad_cam_frequency_appendix}
    \includegraphics[width=0.75\textwidth]{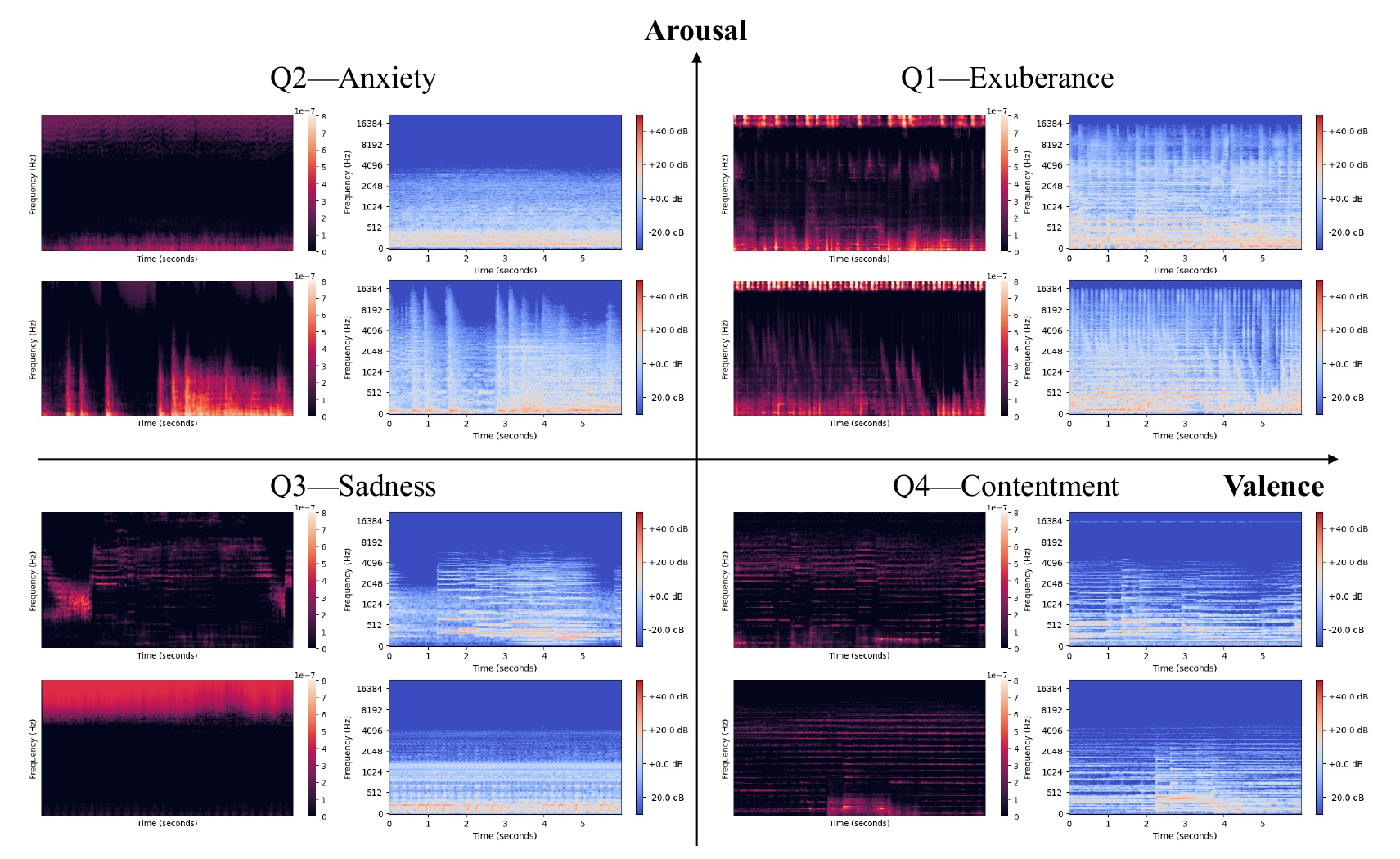}
    \end{center}
     \end{subfigure}
     \hfill
     \begin{subfigure}[b]{\textwidth}
    \begin{center}
        \caption{Time Heatmaps}    
        \label{figure:grad_cam_time_appendix}
    \includegraphics[width=0.75\textwidth]{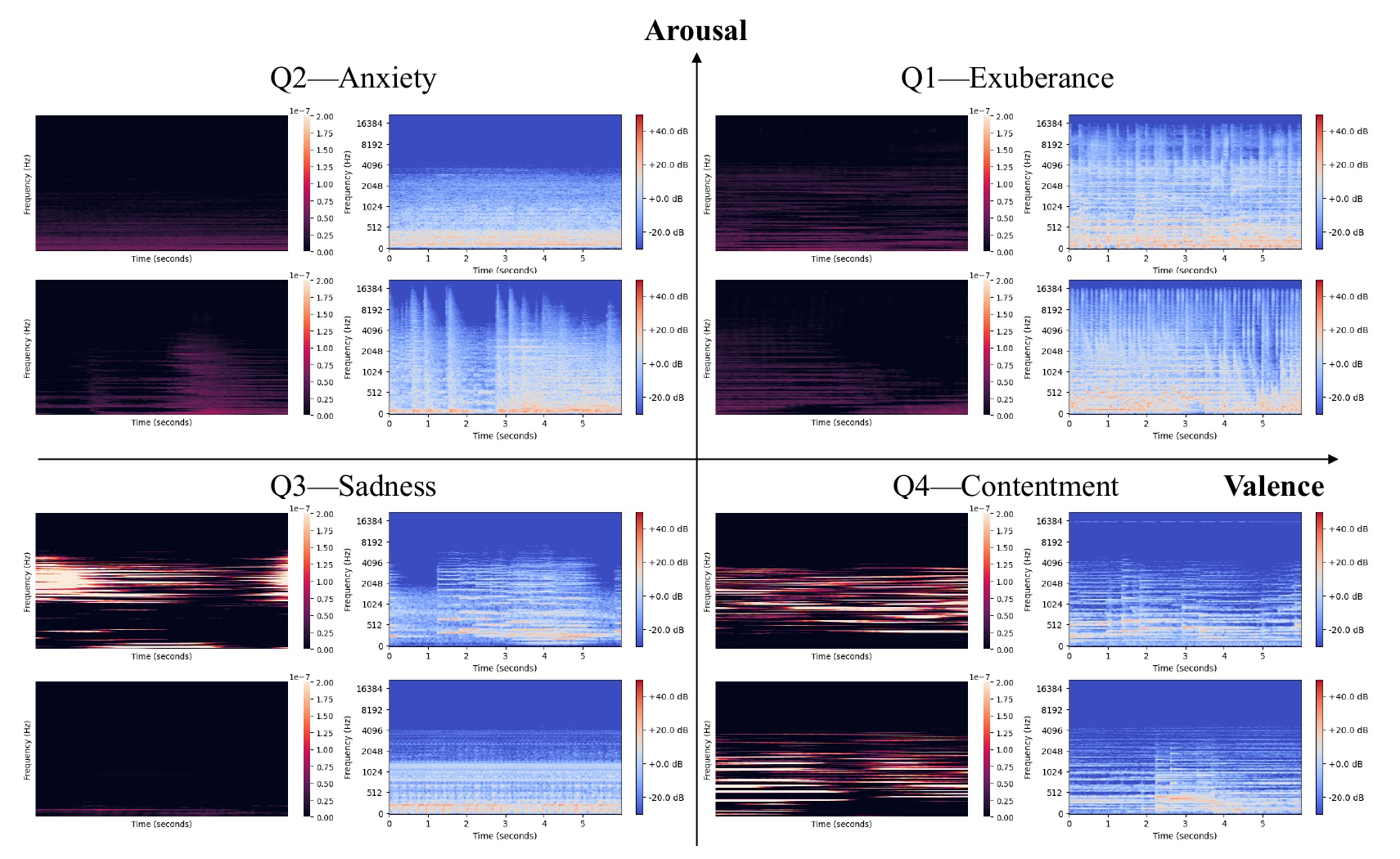}
    \end{center}
     \end{subfigure}
     \\
    \footnotesize
     \raggedright
    Notes: Within each quadrant, the Grad-CAM heatmap (left) corresponds to the spectrogram (right). The heatmap covers the dimensions of the feature map after convolution. The clips are the same as those in Figure 6 of the paper.
\end{figure}

\newpage
\section{Application Study Details}\label{Asection:application}
\setcounter{table}{0}
\setcounter{figure}{0}
Below we provide more details for the ad insertion application. 
\begin{table}[h]
  \footnotesize
  \begin{center}
\caption{Content Video Details}
\begin{tabular}{p{0.13\textwidth}p{0.43\textwidth}p{0.38\textwidth}}
      \hline \hline
      Title & Description & URL \\
      \hline 
        Lost \& Found & Two crocheted stuffed animals try to save each other. Animated and no speech. 6.6 min. & \url{www.youtube.com/watch?v=35i4zTky9pI} \\
        \hline
        Hope & A new hatched turtle learns about its surroundings and tries to get to the ocean. Animated and no speech. 6.2 min. & \url{www.youtube.com/watch?v=1P3ZgLOy-w8}\\
        \hline
        Unspoken & Two people get to know each other and develop a relationship through writing notes. Live-action and some speech. 5.7 min. & \url{www.youtube.com/watch?v=8mpFYQbOCFo}\\
        \hline
        Run With Me & A handicapped high school student participates in the 400m race to prove he doesn't need special treatment. Live-action and speech. 7.9 min. & \url{www.youtube.com/watch?v=EisaD0ZsL3E} \\
      \hline \hline
    \end{tabular}
    \label{table:content_video_details}
    \end{center}
    \vspace{2mm}
    \textit{Note:} The length captures the length of video shown to participants. Some videos are originally longer and we shorten them to start and end at natural times. Table K2 specifies the start and end times.
\end{table}

\begin{table}[htbp]
  \footnotesize
  \centering
  \caption{Ad Insertion Times}
    \begin{tabular}{lcccccccc}
      \hline \hline
      Video & Start & Time 1 & Time 2 & Time 3 & Time 4 & Time 5 & Time 6 & End \\
      \hline 
        Lost \& Found & 0:16 & 1:15 & 2:00 & 3:01 & 3:48 & 5:11 & 6:05 & 6:50 \\
        \hline
        Hope &  0:18 & 1:20 & 2:10 & 2:55 & 3:45 & 4:45 & 5:59 & 6:30 \\
        \hline
        Unspoken & 0:01 & 1:01 & 1:57 & 2:54 & 3:26 & 4:12 & 5:14 & 5:43 \\
        \hline
        Run With Me & 4:30 & 5:37 & 7:44 & 9:00 & 9:45 & 10:45 & 11:40 & 12:24  \\
      \hline \hline
      \multicolumn{9}{l}{\textit{Note:} Times are minute:second and based on time since 0:00 rather than time since Start.}
    \end{tabular}
    \label{table:ad_insertion_times}
\end{table}

\begin{table}[htbp]
  \footnotesize
  \centering
  \caption{Ad Details}
\begin{tabular}{lccl}
      \hline \hline
      Brand & Start Time & End Time & URL \\
      \hline 
        Kit Kat & 0:00 & 0:30 & \url{www.youtube.com/watch?v=4X_e3UWS9aA} \\
        Fragile Childhood & 0:23 & 0:53 & \url{www.youtube.com/watch?v=XwdUXS94yNk}\\
        Eli Lilly--Cymbalta & 0:13 & 0:43 & \url{www.youtube.com/watch?v=Nf6Mm__M5RU}\\
        Calm App & 0:00 & 0:30 & \url{www.youtube.com/watch?v=LWisCdA5rB4} \\
      \hline \hline
    \end{tabular}
    \label{table:ad_details}
\end{table}

\begin{figure}[h]
     \centering
     \caption{Human-Tagged Emotion Distributions of Content Videos}
     \begin{subfigure}[b]{0.45\textwidth}
         \centering
         \includegraphics[width=\textwidth]{Figures_final/content_emotion_distribution_Hope.png}
     \end{subfigure}
     \hfill
     \begin{subfigure}[b]{0.45\textwidth}
         \centering
         \includegraphics[width=\textwidth]{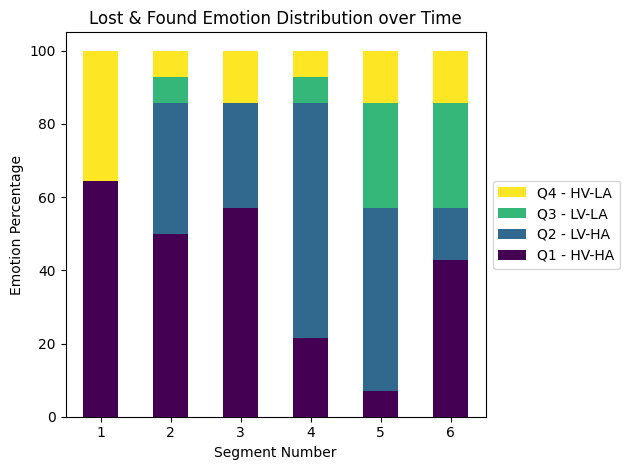}
     \end{subfigure}
     \hfill
     \begin{subfigure}[b]{0.45\textwidth}
         \centering
         \includegraphics[width=\textwidth]{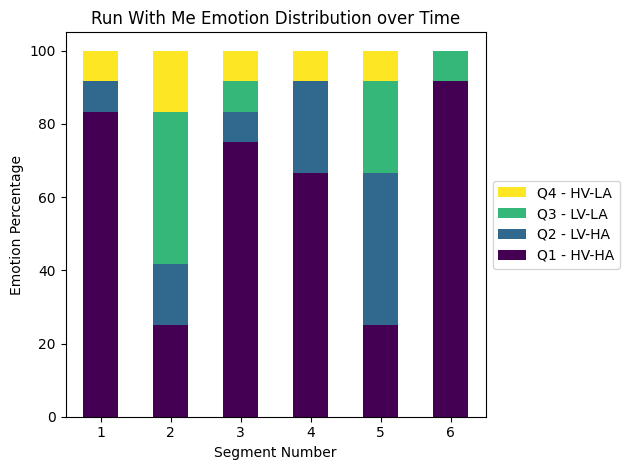}
     \end{subfigure}
     \hfill
     \begin{subfigure}[b]{0.45\textwidth}
         \centering
         \includegraphics[width=\textwidth]{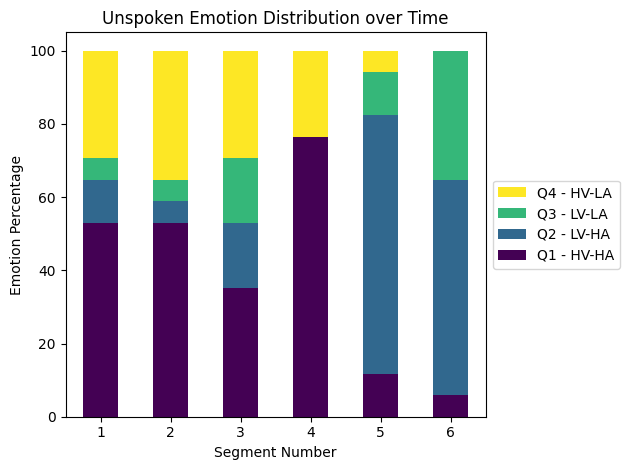}
     \end{subfigure}
    \label{figure:content_video_emotions}
\end{figure}

\begin{figure}
     \centering
     \caption{JS Distance between Ads and Content Videos}
     \begin{subfigure}[b]{0.24\textwidth}
         \centering
         \includegraphics[width=\textwidth]{Figures_final/Hope_Calm_jsd.png}
     \end{subfigure}
     \begin{subfigure}[b]{0.24\textwidth}
         \centering
         \includegraphics[width=\textwidth]{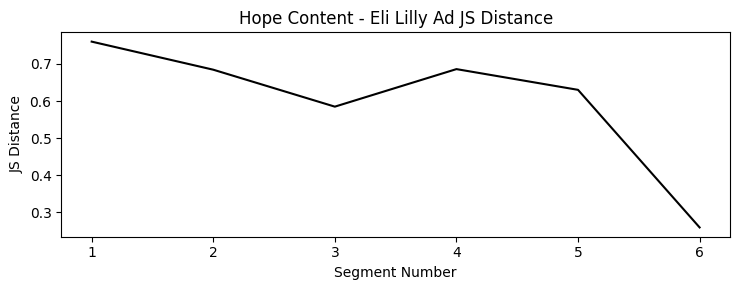}
     \end{subfigure}
     \begin{subfigure}[b]{0.24\textwidth}
         \centering
         \includegraphics[width=\textwidth]{Figures_final/Hope_Fragile_jsd.png}
     \end{subfigure}
     \begin{subfigure}[b]{0.24\textwidth}
         \centering
         \includegraphics[width=\textwidth]{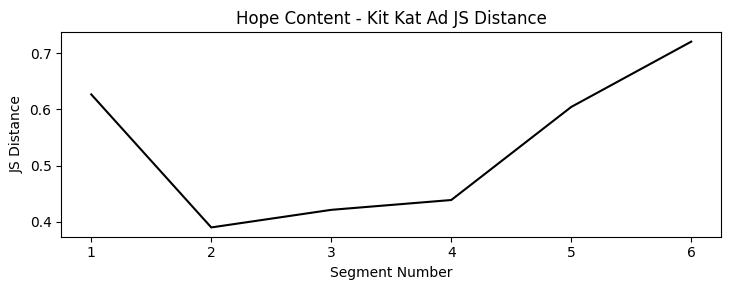}
     \end{subfigure}
     \begin{subfigure}[b]{0.24\textwidth}
         \centering
         \includegraphics[width=\textwidth]{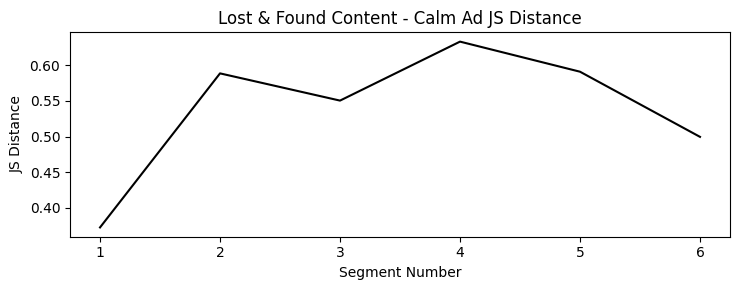}
     \end{subfigure}
     \begin{subfigure}[b]{0.24\textwidth}
         \centering
         \includegraphics[width=\textwidth]{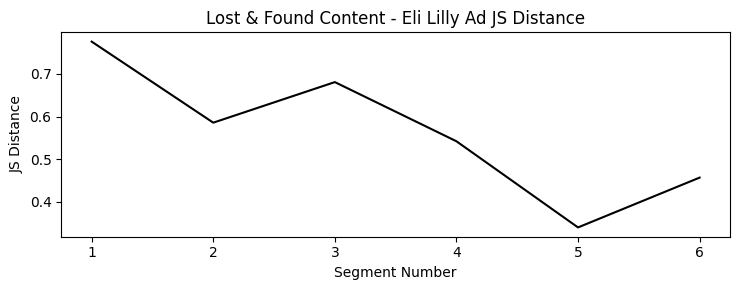}
     \end{subfigure}
     \begin{subfigure}[b]{0.24\textwidth}
         \centering
         \includegraphics[width=\textwidth]{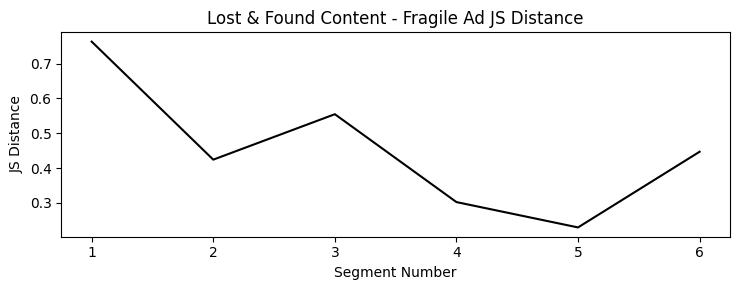}
     \end{subfigure}
     \begin{subfigure}[b]{0.24\textwidth}
         \centering
         \includegraphics[width=\textwidth]{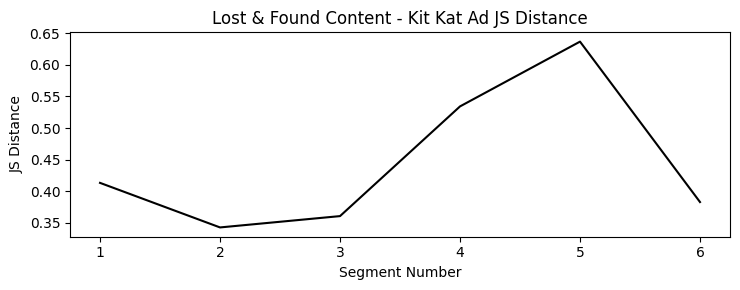}
     \end{subfigure}
     \begin{subfigure}[b]{0.24\textwidth}
         \centering
         \includegraphics[width=\textwidth]{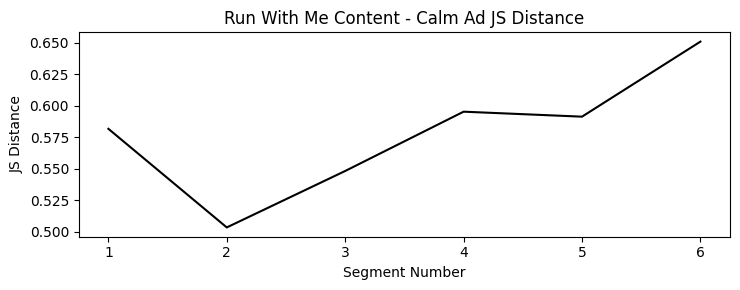}
     \end{subfigure}
     \begin{subfigure}[b]{0.24\textwidth}
         \centering
         \includegraphics[width=\textwidth]{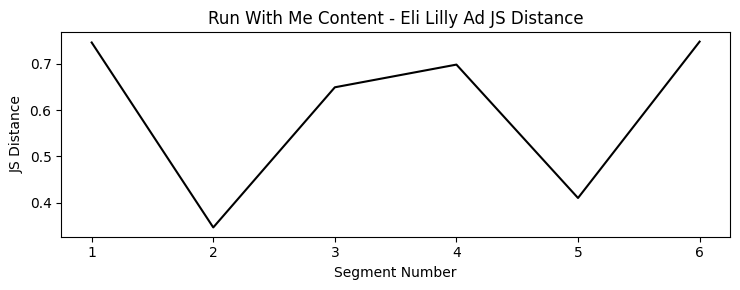}
     \end{subfigure}
     \begin{subfigure}[b]{0.24\textwidth}
         \centering
         \includegraphics[width=\textwidth]{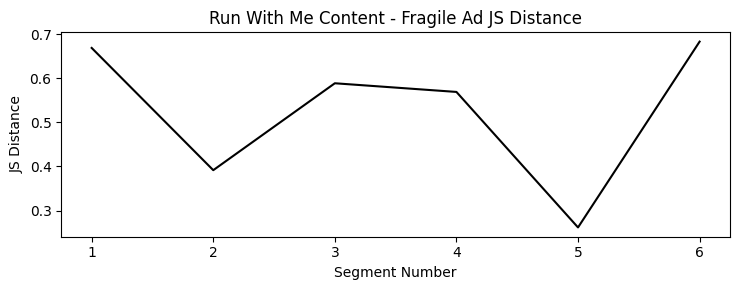}
     \end{subfigure}
     \begin{subfigure}[b]{0.24\textwidth}
         \centering
         \includegraphics[width=\textwidth]{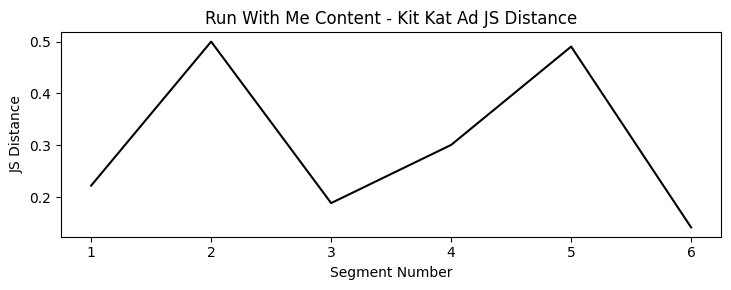}
     \end{subfigure}
     \begin{subfigure}[b]{0.24\textwidth}
         \centering
         \includegraphics[width=\textwidth]{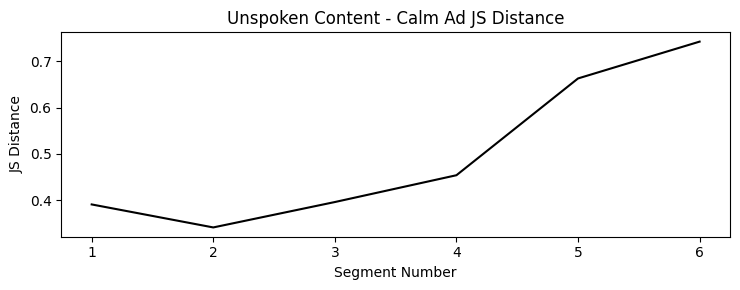}
     \end{subfigure}
     \begin{subfigure}[b]{0.24\textwidth}
         \centering
         \includegraphics[width=\textwidth]{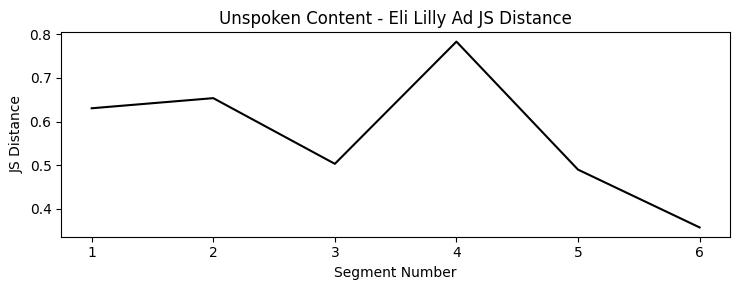}
     \end{subfigure}
     \begin{subfigure}[b]{0.24\textwidth}
         \centering
         \includegraphics[width=\textwidth]{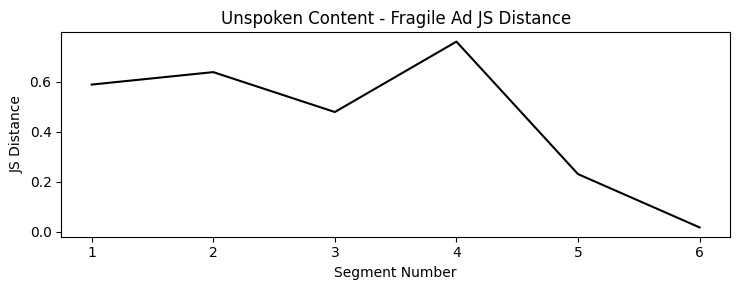}
     \end{subfigure}
     \begin{subfigure}[b]{0.24\textwidth}
         \centering
         \includegraphics[width=\textwidth]{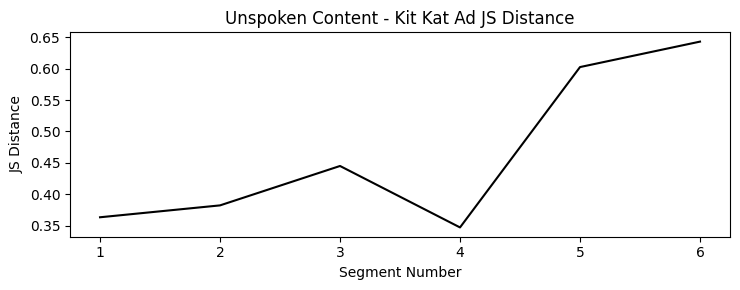}
     \end{subfigure}
    \label{figure:js_distances}
\end{figure}

\subsection{Incorporating Emotion Data from Images and Text}\label{section:ms_azure}

Although music is a key driver of emotion in video, other features like text (what is said), voice tonality (how it is said), and images can also influence emotion. We use Microsoft Azure to try to extract the emotion related to these other features to explore the impact of incorporating emotion data from other modalities beyond the background audio. Azure's Video Indexer emotion detection algorithm predicts emotion from speech text and voice tonality. The possible emotions include joy, fear, anger, and sadness. Azure's Face API detects emotion based on facial expressions from images.\footnote{Starting June 2022, users must apply to use the API due to accuracy concerns for specific demographic groups.} The Face API treats the emotion prediction task as a multiclass problem so the probabilities over the eight possible classes (anger, contempt, disgust, fear, happiness, neutral, sadness, surprise) sum to one.

\begin{figure}[h]
    \centering
    \caption{Content Speech and Facial Emotion Distribution}
    \label{figure:content_face_emotion}
    \subcaption{Speech Emotion - Video Indexer}
    \includegraphics[width=8cm]{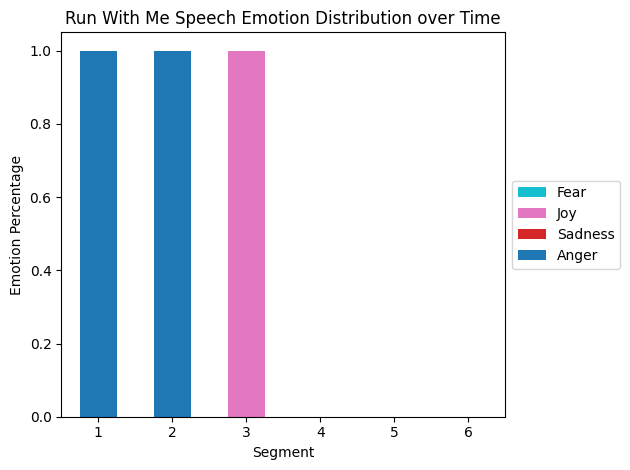}
    \includegraphics[width=8cm]{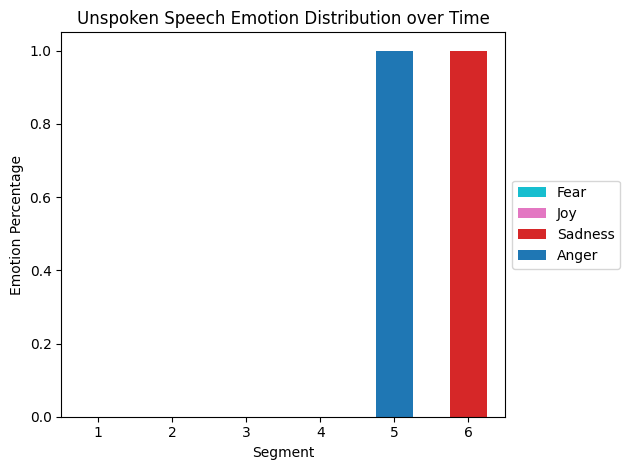}
    \subcaption{Facial Emotion - Face API}
    \includegraphics[width=8cm]{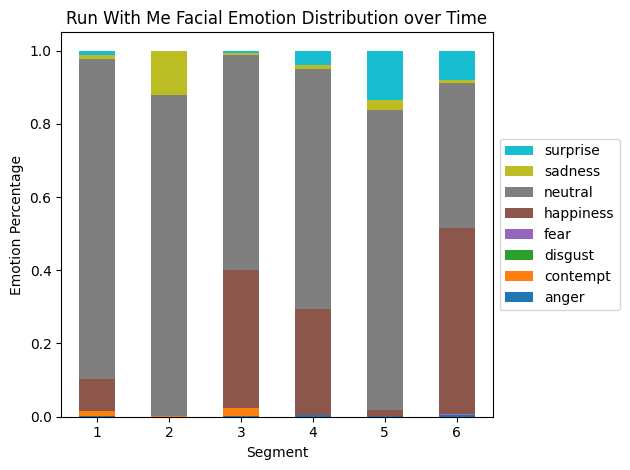}
    \includegraphics[width=8cm]{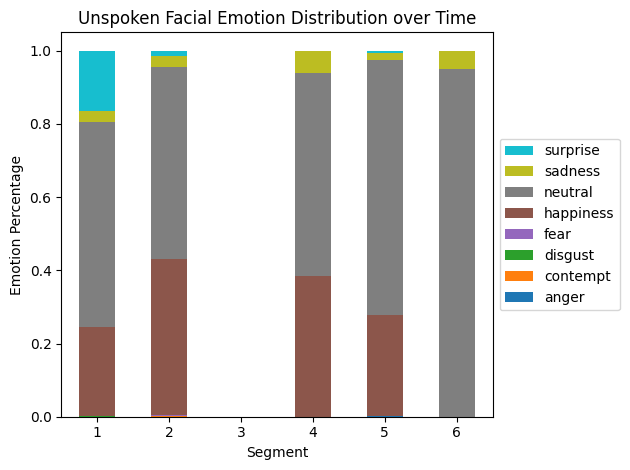}
    \subcaption{Human-tagged Emotion}
    \includegraphics[width=8cm]{Figures_final/content_emotion_distribution_Run.png}
    \includegraphics[width=8cm]{Figures_final/content_emotion_distribution_Unspoken.png}
\end{figure}

For the content videos, the Video Indexer predicted no speech emotion for two of the content videos since they do not contain speech. The first row in Figure \ref{figure:content_face_emotion} shows the distribution of speech emotion for the remaining two videos. Each distribution is defined as the average over the 30 seconds prior to the ad insertion time. The third row shows the human-tagged emotion distributions for comparison. For Run With Me, Segment 3 aligns with the human-tagged emotion but Segments 1 and 2 less so. For Unspoken, Segments 4 and 5 are identified to be negative by the Video Indexer as well as by humans. 

The Face API predicted no emotion for two of the content videos since they do not contain human faces but instead contain animated animal faces. The second row in Figure \ref{figure:content_face_emotion} shows the distribution of facial emotion for the remaining two videos. Each facial emotion distribution is defined as the average over the 30 seconds prior to the ad insertion time. For the most part, the faces are predicted to be either neutral or happy. For Run With Me, we see that there are some similarities with human-tagged emotion in that Segments 3, 4, and 6 are higher in facial happiness and higher in Q1. 

Given the little speech present in the first six seconds of the four ads, we cannot use the Video Indexer for speech emotion in our setting. We can, however, use the Face API in combination with the music emotion classifiers to determine the optimal emotion-based ad insertion point for the two content videos Run With Me and Unspoken. We calculate the JS distance based on face emotion and the JS distance based on music emotion for each ad insertion and ad combination and sum the two distances to determine which ad insertion point is the most emotionally similar for each ad and content video combination. Following the same procedure used with music emotion we calculate the average human-tagged JS distance, skip rate, and recall rate for each model. Table \ref{table:automatic_insertion_face} compares these measures from using face and music emotion versus using only music emotion for the two content videos Unspoken and Run With Me.

\begin{table}[htbp]
  \footnotesize
  \centering
  \caption{Face and Music Emotion vs. Only Music Emotion - Unspoken and Run With Me }
    \begin{tabular}{llccc}
      \hline \hline
      Feature & Model & JS Distance  & Skip Rate & Recall Rate\\
      \hline 
      \textbf{Music Emotion} \T \\
      Mel - Harmonics & CNN & 0.429 & 42.0\% & 48.9\% \\
      Mel - Harmonics + \TempoF & CNN + RF & 0.395 & 40.7\% & 48.8\% \\ 
      \textbf{Face and Music Emotion} \\
      Mel - Harmonics & CNN & 0.459 & 46.1\% & 49.9\% \\ 
      Mel - Harmonics + \TempoF & CNN + RF & 0.445 & 47.6\% & 49.0\% \\ 
      \hline \hline
    \end{tabular}
    \label{table:automatic_insertion_face}
\end{table}

Including face emotion slightly improves the recall rate but hurts the skip rate. Overall, there is potential in incorporating emotion information from images and text but the existing tools are limited in their ability to extract emotion information from short clips (i.e., first six seconds of ads) and animated videos. The results suggest that audio models like the one included in Azure's Video Indexer service could benefit from incorporating music emotion classification.

\end{APPENDIX}

\end{document}